\documentclass[10pt,twocolumn]{article}

\usepackage[utf8]{inputenc}
\usepackage[T1]{fontenc}
\usepackage{amsmath,amssymb,amsfonts,amsthm}
\usepackage{bm}
\usepackage{graphicx}
\graphicspath{{figures/}{./}}
\usepackage{booktabs}
\usepackage{geometry}
\usepackage{siunitx}
\usepackage{xcolor}
\usepackage{hyperref}
\usepackage{url}
\usepackage{enumitem}
\usepackage{balance}
\hypersetup{colorlinks=true,linkcolor=blue,citecolor=blue,urlcolor=blue}

\newcommand{\R}{\mathbf{R}}
\newcommand{\rvec}{\mathbf{r}}
\newcommand{\pvec}{\mathbf{p}}
\newcommand{\muvec}{\bm{\mu}}

\newcommand{\braket}[2]{\langle #1 \vert #2\rangle}
\newcommand{\mel}[3]{\langle #1 \vert #2 \vert #3\rangle}
\newcommand{\Eh}{E_{\mathrm{h}}}

\newcommand{\Tr}{\operatorname{Tr}}
\newcommand{\diag}{\operatorname{diag}}
\newcommand{\OrbO}{\textsf{OrbACE-O}}
\newcommand{\OrbA}{\textsf{OrbACE-A}}

\theoremstyle{plain}
\newtheorem{proposition}{Proposition}
\theoremstyle{remark}
\newtheorem{remark}{Remark}

\title{\textbf{OrbACE: Orbital and Atomic Cluster Expansion}\\[0.35em]
\large Equivariant Message Passing on Electronic-Structure Graphs}

\author{Etinosa Osaro\,$^{*}$ and Dario Rocca\,$^{*}$ \\[0.35em]
\normalsize PsiQuantum, 700 Hansen Way, Palo Alto, CA 94304 \\[0.25em]
\normalsize $^{*}$Corresponding authors:
\href{mailto:eosaro@psiquantum.com}{eosaro@psiquantum.com},
\href{mailto:drocca@psiquantum.com}{drocca@psiquantum.com}}
\date{\today}

\begin{document}
\onecolumn
\maketitle

\begin{abstract}
Accurate electronic-structure reference data are expensive, motivating
interatomic potentials that learn effectively from limited training sets.
We introduce \textbf{OrbACE} (Orbital and Atomic Cluster Expansion), a
framework that supplements an $E(3)$-equivariant cluster-expansion backbone
with inexpensive electronic-structure descriptors. On ANI-1x at the
$\omega$B97X/6-31G$^*$ level, its atom-centered variant, \OrbA, trained on
10,000 configurations and evaluated on 442,000 held-out configurations,
achieves energy and force mean absolute errors of 3.16\,meV/atom and
76\,meV/{\AA} in direct learning, compared with 9.61\,meV/atom and
146\,meV/{\AA} for MACE with matched backbone settings. This reduces the
energy error by a factor of three and approximately halves the force
error. Its energy error is also below that of the best delta-corrected
MACE baseline, 5.00\,meV/atom, although its force error is slightly higher.
Combining \OrbA\ with a Hartree--Fock delta baseline further reduces the
errors to 1.48\,meV/atom and 35\,meV/{\AA}.

\OrbA\ builds atom- and pair-resolved invariants from a non-self-consistent
extended-H\"uckel model evaluated within the differentiable forward pass,
yielding forces consistent with the predicted energy. The orbital
variant, \OrbO, instead uses localized Hartree--Fock orbitals as graph
nodes, with scalar and equivariant electronic descriptors. Its forces
are evaluated at fixed descriptors, omitting their response to nuclear
motion; in direct learning on ANI-1x, its energy error is approximately
three times that of \OrbA. The three benchmarks considered---ANI-1x,
revised MD17, and 3BPA---cover neutral, closed-shell, non-periodic
molecules containing H, C, N, and O. Performance outside this domain
remains to be established.
\end{abstract}
\twocolumn

\section{Introduction}
\label{sec:intro}

Accurate reference data are expensive, so potentials that learn well from
limited training sets are valuable. Machine-learned interatomic potentials approximate the map from a nuclear
configuration to its Born--Oppenheimer energy,
$(\{Z_A\},\{\R_A\}) \mapsto E$, and to the forces $-\partial E/\partial\R_A$.
From high-dimensional neural-network potentials~\cite{bpnn} and kernel models
on smooth density descriptors~\cite{soap} to equivariant message
passing~\cite{nequip}, the Atomic Cluster Expansion (ACE)~\cite{ace} and its
message-passing realization MACE~\cite{mace}, these models now reach chemical
accuracy at a cost orders of magnitude below the method that generated their
labels. Their input, however, is only atomic numbers and positions: bond order,
hybridization, charge transfer and polarization must be reconstructed from
data, although quantum chemistry computes them directly, and in small Gaussian
bases cheaply.

A central motivation is the regime in which accurate reference labels are scarce: high-level correlated electronic-structure calculations, and potentially future quantum-computing calculations, may provide energies for only a limited set of configurations. Combining electronic-structure descriptors with multi-fidelity learning offers a route to exploit abundant inexpensive calculations alongside these sparse high-fidelity labels. Here we investigate this strategy using density-functional reference data; extending it to correlated classical or quantum-computing reference energies is a prospective application.

OrbACE evaluates an inexpensive electronic-structure model first and hands the
result to the network as node and edge features. The premise is shared with
OrbNet~\cite{orbnet}, which predicts density-functional energies from blocks of
semi-empirical Fock, overlap and density matrices.
OrbNet Denali~\cite{orbnetdenali} scales that approach to 2.3 million
density-functional calculations across seventeen elements, and
OrbNet-Equi~\cite{orbnetequi}, through its UNiTE network, learns equivariantly
from the tight-binding matrices. A related line of work runs in the opposite
direction and predicts the electronic structure itself: SchNOrb~\cite{schnorb}
and PhiSNet~\cite{phisnet} predict the Hamiltonian and wavefunction in an
atomic-orbital basis, the latter equivariantly. Another keeps atoms as the only
nodes but gives the potential electronic degrees of freedom, as
SpookyNet~\cite{spookynet} does for total charge and spin and
AIMNet2~\cite{aimnet2} does across neutral and charged organic and
element-organic chemistry. What is new here is
the combination with a body-ordered cluster expansion, equivariant descriptors
resolved on localized orbitals as graph nodes, and the requirement that the
energy be differentiated exactly, which is what makes the featurizer usable as a force
field rather than only as an energy model. The idea is instantiated in two
ways, which differ in what a graph node is.

In the \emph{orbital} cluster expansion, \OrbO, the nodes are localized
molecular orbitals. A Hartree--Fock (HF) calculation in a small Gaussian basis
is followed by a localization, and each orbital becomes a node carrying its
energy, occupancy, integral diagonals, spatial extent, shape invariants, and
equivariant polarization and quadrupole tensors; edges carry Fock, Coulomb,
exchange, core-Hamiltonian and overlap couplings and the energy denominator
that governs orbital mixing. Bonds and lone pairs are first-class objects, at
the price of a self-consistent-field (SCF) solution and a localization per
configuration.

In the \emph{atomic} cluster expansion, \OrbA, the nodes are atoms and the
descriptors come from a one-shot, non-self-consistent extended-H\"uckel model
evaluated in closed form inside the forward pass, so the energy is an explicit
function of the coordinates and reverse-mode differentiation returns exact
forces. The representation is coarser, atom- rather than orbital-resolved, but
strictly conservative, and costs one small dense eigendecomposition.

The two models share a backbone, an energy decomposition, a short-range
repulsion, a training objective, and the delta-learning and multi-fidelity
machinery built on top (Sec.~\ref{sec:backbone}). Two problems shape the rest
of the design. The first is differentiability: with precomputed features, as in
\OrbO, the energy reaches the coordinates only through the construction that
places nodes relative to nuclei (Sec.~\ref{sec:orbpos}); 
with features recomputed in the forward pass, as in \OrbA, automatic differentiation can include their dependence on nuclear coordinates, but obtaining correct forces requires special treatment of eigenvalue degeneracies, where standard eigenvector derivatives become singular (Sec.~\ref{sec:spectral}). The second is scale heterogeneity: 
orbital energies span core and valence energy scales, spatial second moments have different units and characteristic magnitudes, and occupations are discrete.
\OrbO\ normalizes per group and per sample and \OrbA\ against fixed dataset
statistics, because some of \OrbA's features encode an absolute energy whose
magnitude a per-sample normalization would destroy (Secs.~\ref{sec:embed},
\ref{sec:atom-embed}). Sections~\ref{sec:background}--\ref{sec:training}
develop the method and Sec.~\ref{sec:results} reports the benchmarks;
implementation detail is collected in the Supporting Information
(Secs.~\ref{si:notation}--\ref{si:results}).


\section{Background and notation}
\label{sec:background}

\subsection{Hartree--Fock, molecular orbitals, and density matrices}

In the Hartree--Fock approximation the many-electron wavefunction is a single
Slater determinant of molecular orbitals (MOs) $\{\psi_i\}$, each expanded in a
finite atom-centered Gaussian basis $\{\chi_\mu\}$,
\begin{equation}
  \psi_i(\rvec) = \sum_{\mu=1}^{n_{\mathrm{AO}}} C_{\mu i}\,\chi_\mu(\rvec),
  \qquad
  \mathbf{C}\in\mathbb{R}^{n_{\mathrm{AO}}\times n_{\mathrm{MO}}} .
\end{equation}
where $n_{\mathrm{AO}}$ is the number of atom-centered basis functions and
$n_{\mathrm{MO}}$ the number of molecular orbitals. The coefficients solve the
Roothaan--Hall generalized eigenproblem
\begin{equation}
  \mathbf{F}\mathbf{C}=\mathbf{S}\mathbf{C}\bm{\varepsilon},
  \qquad
  \mathbf{S}_{\mu\nu}=\braket{\chi_\mu}{\chi_\nu},
  \label{eq:roothaan}
\end{equation}
where $\mathbf{S}$ is the atomic-orbital (AO) overlap,
$\bm{\varepsilon}=\diag(\varepsilon_i)$ the canonical orbital energies, and
\begin{equation}
  \mathbf{F} = \mathbf{h} + \mathbf{v}_{\mathrm{HF}}[\mathbf{P}],
  \qquad
  \mathbf{h} = \mathbf{T} + \mathbf{V}_{\mathrm{nuc}},
  \label{eq:fock}
\end{equation}
the Fock matrix, which adds to the core Hamiltonian $\mathbf{h}$ (kinetic energy
plus nuclear attraction) the Coulomb and exchange potential of the one-particle
density matrix. The one-particle and energy-weighted density matrices are
\begin{equation}
  \mathbf{P} = \sum_i n_i\,\mathbf{c}_i \mathbf{c}_i^\top,
  \qquad
  \mathbf{W} = \sum_i n_i\,\varepsilon_i\,\mathbf{c}_i \mathbf{c}_i^\top ,
  \label{eq:densmat}
\end{equation}
with $\mathbf{c}_i$ the $i$-th column of $\mathbf{C}$ and $n_i\in\{0,1,2\}$ its
occupancy. Equations~\eqref{eq:roothaan}--\eqref{eq:fock} are solved by
self-consistent iteration. Hartree--Fock treats electrostatics, Pauli exchange
and orbital structure at the mean-field level and omits dynamic correlation,
which a featurizer does not need. $\mathbf{P}$ and $\mathbf{W}$ are invariant
under rotations among orbitals of equal occupancy and smooth in the geometry
away from level crossings.

\subsection{Equivariance conventions}

A function $f$ of a configuration is \emph{invariant} if
$f(\mathbf{Q}\R)=f(\R)$ for every $\mathbf{Q}\in O(3)$, and
\emph{equivariant} of degree $\ell$ if
$f(\mathbf{Q}\R)=\mathbf{D}^{(\ell)}(\mathbf{Q})f(\R)$, where $\mathbf{D}^{(\ell)}(\mathbf{Q})$ is the Wigner $D$-matrix for rotations, extended to improper rotations with the appropriate parity. We use the $e3nn$
convention~\cite{e3nn}, in which an irrep is written $\ell e$ or $\ell o$
according to its parity and a feature space as a direct sum with multiplicities,
e.g., $64\times 0e + 64\times 1o + 64\times 2e$. Scalars are $0e$, a position
difference is $1o$, and a symmetric traceless rank-2 tensor is $2e$. Energy is
$0e$, and forces are obtained by differentiation, which guarantees their $1o$
character; the corresponding requirement on the electronic descriptors is
established in Sec.~\ref{sec:equivariance}. Electronic quantities are kept in
atomic units, geometry in {\AA}ngstr\"om, and labels in eV and eV/{\AA};
uppercase Latin indices run over atoms, lowercase over orbitals, and Greek over
atomic orbitals (Sec.~\ref{si:notation}).

\section{The shared equivariant backbone}
\label{sec:backbone}

After featurization both models produce graphs of the same form, with
irreps-valued node features and edges carrying a displacement vector and invariant chemical
scalars; this section describes the machinery that acts on it.

\subsection{What a node is, in each model}

\begin{figure*}[!tbp]
\centering
\includegraphics[width=\linewidth]{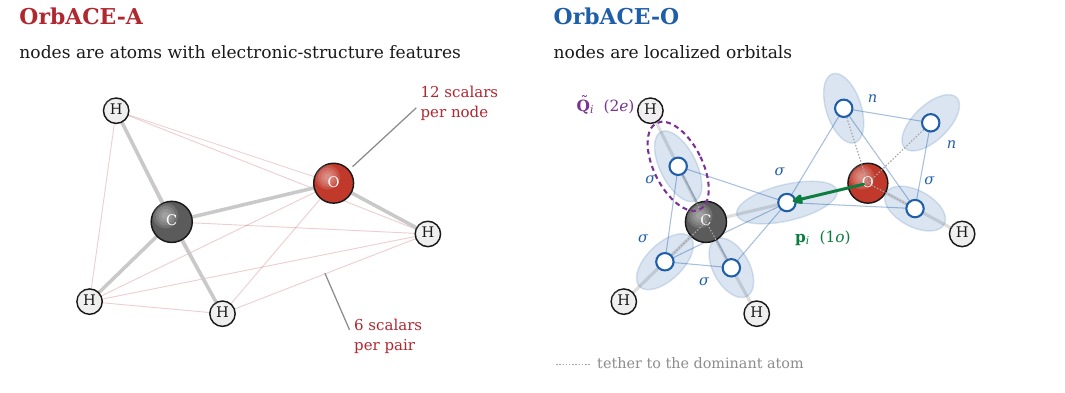}
\caption{What carries a graph node in each model, drawn on methanol.
\OrbA\ places one node on each nucleus, described by the twelve invariants of
Table~\ref{tab:tbnode}, with six more on each atom pair. \OrbO\ places one node
on each localized molecular orbital, at its centroid, so that nodes sit on
bonds ($\sigma$) and lone pairs ($n$). Each orbital carries thirty invariant
scalars together with the polarization vector $\pvec_i$ of
Eq.~\eqref{eq:pol}, drawn from the dominant nucleus to the centroid, and the
traceless quadrupole $\tilde{\mathbf{Q}}_i$ of Eq.~\eqref{eq:quadtl}, sketched
as an ellipse; each tensor is annotated on one orbital only. Dotted tethers
mark the dominant atom of Eq.~\eqref{eq:anchor}; grey sticks are the molecular
skeleton, thin coloured lines are graph edges, and core orbitals are omitted.}
\label{fig:representations}
\end{figure*}

The single structural difference between the two models is summarized in
Table~\ref{tab:variants} and drawn in Fig.~\ref{fig:representations}.

\begin{table*}[!tbp]
\centering
\caption{The two OrbACE models. The backbone, energy decomposition,
short-range repulsion, training objective, and delta-learning machinery are
shared; the representation layer is not.}
\label{tab:variants}
\vspace{3pt}
\small
\renewcommand{\arraystretch}{1.2}
\begin{tabular}{p{0.21\linewidth}p{0.35\linewidth}p{0.35\linewidth}}
\toprule
 & \textbf{\OrbO} (Sec.~\ref{sec:orbital}) & \textbf{\OrbA} (Sec.~\ref{sec:atomic}) \\
\midrule
Graph nodes & Localized molecular orbitals & Atoms \\
Node positions & Mulliken-weighted, dominant-atom-anchored orbital centroids & Nuclear coordinates \\
Electronic model & Hartree--Fock, self-consistent, in a Gaussian basis & Extended H\"uckel, non-self-consistent, minimal valence basis \\
When it is evaluated & Once per configuration, ahead of training & Inside every forward pass \\
Node features & 10 scalar QC invariants, 15-dim shell encoding, 5-dim per-$\ell$ coefficient norms, equivariant $1o$ and $2e$ tensors & 12 invariant scalars from the density and Hamiltonian matrices \\
Edge features & 6 MO-pair couplings $F_{ij},J_{ij},K_{ij},h_{ij},|\Delta\varepsilon_{ij}|,S_{ij}$ & 6 atom-pair invariants: Mayer bond order, four block norms, one energy coupling \\
Feature normalization & Per-group, per-sample \texttt{LayerNorm} & Dataset-level standardization \\
Source of equivariance & Geometry \emph{and} equivariant electronic tensors & Geometry alone \\
Status of the forces & Gradient of the energy at fixed features & Exact gradient of the energy \\
Additional terms & Occupation gate; atom-resolved correction; optional electrostatics to dipole order & Optional electrostatics, monopole only \\
\bottomrule
\end{tabular}
\end{table*}

\subsection{What the backbone does}
\label{sec:backbone-summary}

The backbone is that of MACE~\cite{mace}, used unmodified; the equations are in
Sec.~\ref{si:backbone}. Each edge carries a Bessel radial basis multiplied by a
polynomial envelope that vanishes with its first two derivatives at the cutoff,
and the real spherical harmonics of the edge direction to $\ell_{\max}=2$.
Invariant chemical edge features are projected into the radial space and
added, under the same envelope, so that a coupling is switched off smoothly at
the cutoff rather than truncated. Each of the $L$ layers applies an equivariant
convolution followed by a symmetric product that forms tensor powers of the
message up to correlation order $\nu$. That product is what makes the model a
cluster expansion: a single message is a two-body object, so $\nu=3$ resolves
four-body terms within one layer. Every layer reads out a per-node scalar, and
their sum, rescaled by a training-set force statistic held constant in the
backward pass, is the per-node interaction energy $E^{\mathrm{int}}$.
\label{sec:energy-shared}

The total energy is
\begin{equation}
  E \;=\; \underbrace{E_0}_{\text{species only}}
    \;+\; \underbrace{\sum_{\text{nodes}} E^{\mathrm{int}}}_{E_{\mathrm{inter}}}
    \;+\; E_{\mathrm{corr}}
    \;+\; E_{\mathrm{LR}}
    \;+\; E_{\mathrm{ZBL}} .
  \label{eq:energy-shared}
\end{equation}
The first term is a frozen, geometry-independent baseline counting the nodes of
each species. It absorbs the isolated-atom energies, so that the network fits a
cohesive energy of $O(1\text{--}10)$ eV rather than a total of $O(10^3)$ eV, and
drops out of the forces exactly. \OrbO\ carries a per-orbital-species and a
per-element constant (Sec.~\ref{sec:e0-orb}), \OrbA\ the per-element one only
(Sec.~\ref{sec:atom-energy}). The correction $E_{\mathrm{corr}}$ is specific to
\OrbO\ (Sec.~\ref{sec:orb-readout}), and the electrostatic term
$E_{\mathrm{LR}}$ is disabled in every run reported here
(Sec.~\ref{sec:longrange}). The last term, $E_{\mathrm{ZBL}}$, is a fixed screened-Coulomb repulsion
of Ziegler--Biersack--Littmark form~\cite{zbl} on atom pairs
(Sec.~\ref{sec:zbl}), supplying the short-range wall that equilibrium training
sets sample only sparsely; it is switched off smoothly beyond the sum of
covalent radii.

\subsection{Forces, virials, and stress}
\label{sec:forces-shared}

\begin{figure*}[!tbp]
\centering
\includegraphics[width=\linewidth]{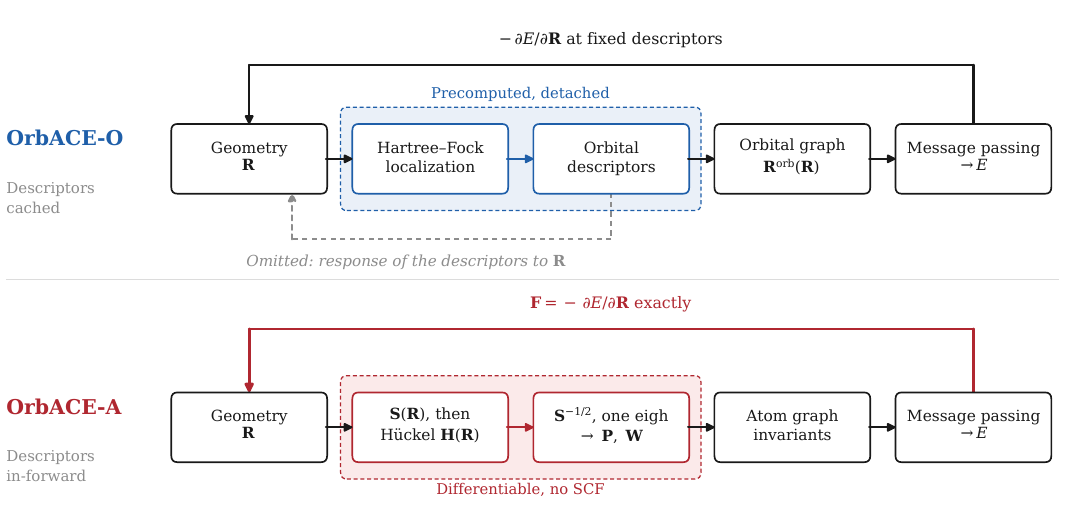}
\caption{The two pipelines, and where the gradient can and cannot reach. Data
flows left to right; the black and red arrows running right to left are
gradient paths. In \OrbO\ the electronic descriptors are computed once and
cached, so the gradient returns to the geometry only through the
orbital-position construction of Eqs.~\eqref{eq:mw-centroid}
and~\eqref{eq:anchor}. The dashed path is the response of the descriptors to
the geometry, the term omitted from Eq.~\eqref{eq:orbforce-split}. In \OrbA\
the overlap, the H\"uckel Hamiltonian, and the density matrices are rebuilt in
every forward pass with nothing detached, so the gradient traverses the entire
chain and is exact (Proposition~\ref{prop:conservative}).}
\label{fig:gradient}
\end{figure*}

Forces are obtained by reverse-mode differentiation of the
geometry-dependent part of Eq.~\eqref{eq:energy-shared},
\begin{equation}
  \mathbf{F}_A = -\frac{\partial}{\partial \R_A}
    \Big(E_{\mathrm{inter}} + E_{\mathrm{corr}}
         + E_{\mathrm{LR}} + E_{\mathrm{ZBL}}\Big),
  \label{eq:forces-shared}
\end{equation}
never by a separate force head: an equivariant network can output a vector per
atom, but such a field is not constrained to be curl-free, so it need not be the
gradient of any energy. For periodic systems, none of which are benchmarked here, the stress follows
from the same backward pass under a symmetric strain $\bm{\varepsilon}$ applied
to positions and cell,
\begin{equation}
  \bm{\sigma} = \frac{1}{V}\,
                \frac{\partial E}{\partial \bm{\varepsilon}}
                \bigg|_{\bm{\varepsilon}=0}.
  \label{eq:stress}
\end{equation}

\begin{remark}
Equation~\eqref{eq:forces-shared} states what is differentiated, not that the
result is the exact gradient of the model's own energy. That depends on whether
the electronic features are themselves functions of $\R$ inside the graph. In
\OrbA\ they are, and the identity is exact (Sec.~\ref{sec:conservative}). In
\OrbO\ they are precomputed constants, and Eq.~\eqref{eq:forces-shared} is the
gradient at fixed features (Sec.~\ref{sec:orb-forces}).
Figure~\ref{fig:gradient} shows the difference and doubles as a roadmap for
Secs.~\ref{sec:orbital} and~\ref{sec:atomic}.
\end{remark}


\section{\OrbO: the orbital cluster expansion}
\label{sec:orbital}

\OrbO\ moves the graph off the nuclei: its nodes are localized molecular
orbitals, its edges orbital pairs, and its features the matrix elements a
quantum chemist would compute to reason about the same system. The
Hartree--Fock calculation, the localization and the featurization are performed
once per configuration and cached; graph construction, message passing and
differentiation run on every training step (upper row of
Fig.~\ref{fig:gradient}).

\subsection{What the featurizer produces}
\label{sec:orb-featurizer}

A restricted Hartree--Fock calculation in a small Gaussian basis, STO-3G by
default, is followed by a Foster--Boys or Pipek--Mezey localization of each
occupancy block (Sec.~\ref{si:orbital}). Localization is what makes a shared
vocabulary of orbital types possible: canonical orbitals are delocalized, a
symmetry-adapted $\pi$ orbital of benzene resembles nothing in ethane, and they
change character discontinuously wherever two eigenvalues cross, whereas
localized orbitals recur with nearly identical descriptors across molecules.
A localized orbital is not a Fock eigenvector, so its energy is taken as the
Fock diagonal in the localized basis, and the off-diagonal elements become the
couplings the graph carries on its edges.

Each orbital is labelled by an integer species encoding its dominant element,
angular momentum and occupancy class, the analogue of the atomic number, and
carries thirty scalar descriptors: ten quantum-chemical invariants
(Table~\ref{tab:orbscalars}), a fifteen-dimensional binary encoding of the
contributing atomic-orbital shells, and the squared coefficient mass in each
angular channel. One of the ten, the orbital partition energy
$\tfrac12 n_i(h_{ii}+F_{ii})$, sums exactly to the Hartree--Fock electronic
energy, but per-orbital normalization removes that absolute scale before the
network sees it. Each orbital pair that forms an edge carries six couplings: the
Fock, Coulomb, exchange, core-Hamiltonian and overlap matrix elements, and the
energy denominator $|\varepsilon_i-\varepsilon_j|$ that governs how strongly two
orbitals mix (Sec.~\ref{si:orbital}).

\subsection{Mulliken weights and orbital centroids}
\label{sec:mulliken}

Two constructions relate orbitals back to atoms. The Mulliken gross
population~\cite{mulliken} of orbital $i$ on atom $A$ is normalized to a weight
\begin{equation}
  w_{iA} \;=\;
  \frac{\sum_{\mu\in A} C_{\mu i}\,(\mathbf{S}\mathbf{C})_{\mu i}}
       {\sum_{B}\sum_{\mu\in B} C_{\mu i}\,
        (\mathbf{S}\mathbf{C})_{\mu i}},
  \qquad \sum_A w_{iA} = 1 ,
  \label{eq:mulliken}
\end{equation}
where $\mu\in A$ runs over the AOs centered on $A$. Individual weights can be
negative in a minimal basis, but it is the \emph{signed} sum that must be one:
that makes the orbital position of Eq.~\eqref{eq:mw-centroid} an affine
combination of nuclear positions, which translates with the molecule. The
weights fix the dominant atom and hence the species (Eq.~\ref{eq:species}),
define a differentiable orbital position (Eq.~\ref{eq:mw-centroid}), and pool
per-orbital scalars back onto atoms in the readout (Eq.~\ref{eq:atompool}).
The orbital centroid is the position expectation value,
\begin{equation}
  \rvec_i = \mel{\psi_i}{\hat{\rvec}}{\psi_i}
          = \sum_{\mu\nu} C_{\mu i}\,\mathbf{d}_{\mu\nu}\,C_{\nu i},
  \qquad
  \mathbf{d}_{\mu\nu} = \mel{\chi_\mu}{\hat{\rvec}}{\chi_\nu}.
  \label{eq:centroid}
\end{equation}

\subsection{Equivariant node features}
\label{sec:equiv-feats}

A purely invariant node description would force the network to recover all
orientational information from the geometry of the orbital graph. \OrbO\
supplies two lab-frame electronic quantities that transform as proper irreps:
\begin{align}
  \pvec_i &= \rvec_i - \R_{A^{\mathrm{dom}}_i}
    && (\ell{=}1,\ \text{polarization},\ a_0), \label{eq:pol}\\
  \tilde{\mathbf{Q}}_i &= \mathbf{Q}_i
    - \tfrac13 \big(\Tr \mathbf{Q}_i\big)\,\mathbf{I}
    && (\ell{=}2,\ \text{traceless quadrupole},\ a_0^2). \label{eq:quadtl}
\end{align}
The polarization points from the dominant nucleus to the centroid, encoding
where a lone pair projects or a bond is polarized; the traceless quadrupole
carries the orientational content that the quadrupole eigenvalues discard.
Each is mapped into the $e3nn$ spherical basis by a fixed change of basis,
rescaled by a fixed constant, and mixed by an equivariant linear layer into
eight channels each of $1o$ and $2e$, so the initial node irreps are
$m\times 0e + 8\times 1o + 8\times 2e$. Only equivariant operations act on
them, and both are precomputed constants that carry no gradient
(Sec.~\ref{sec:orb-forces}).

\subsection{Orbital positions and dominant-atom anchoring}
\label{sec:orbpos}

In the forward pass each orbital position must be an explicit function of the
nuclear coordinates, or no force can flow from the graph back to the geometry.
\OrbO\ therefore rebuilds a \emph{Mulliken-weighted centroid},
\begin{equation}
  \tilde{\rvec}_i = \sum_A w_{iA}\,\R_A ,
  \label{eq:mw-centroid}
\end{equation}
with the frozen weights of Eq.~\eqref{eq:mulliken}. This linearizes the
centroid of Eq.~\eqref{eq:centroid} in the nuclear positions at fixed
electronic structure; it is exact when the orbital moves rigidly with its atoms
and differentiable everywhere. On its own it can place two orbitals at nearly
the same point, since the weighted mean of a delocalized orbital's atoms can
land anywhere, and coincident nodes are catastrophic for a distance-based graph.
Orbitals are therefore \emph{anchored} to their dominant atom,
\begin{equation}
  \R_i^{\mathrm{orb}} = \R_{A^{\mathrm{dom}}_i}
    + s_{\mathrm{spread}}\,
      \big(\tilde{\rvec}_i - \R_{A^{\mathrm{dom}}_i}\big),
  \label{eq:anchor}
\end{equation}
with $s_{\mathrm{spread}}$ a global learnable scalar initialized at $1$, so
that orbitals on different atoms are separated on the scale of a bond length.
The dominant atom is the largest $|w_{iA}|$, and near-ties are common, so an
arbitrarily small change in the weights can switch it, moving the node and
making the predicted energy discontinuous there (Sec.~\ref{sec:orbgraph}).

\subsection{The force path and its status}
\label{sec:orb-forces}

Forces follow Eq.~\eqref{eq:forces-shared} through the chain
\begin{equation}
  \begin{gathered}
  \R_A
  \;\longrightarrow\; \tilde{\rvec}_i \;\text{(Eq.~\ref{eq:mw-centroid})}
  \;\longrightarrow\; \R^{\mathrm{orb}}_i \;\text{(Eq.~\ref{eq:anchor})}
  \\ \;\longrightarrow\; \mathbf{v}_{ij}
  \;\longrightarrow\; Y_\ell,\ b_n
  \;\longrightarrow\; \mathbf{h}_i
  \;\longrightarrow\; E .
  \end{gathered}
\end{equation}
Moving a nucleus moves every orbital anchored to it or populating it, and
hence the edge vectors, the harmonics, and the messages. The electronic
features themselves, meaning the scalar and edge integrals, the polarization,
the quadrupole, and the edge list, are precomputed and carry no gradient. The
reported force is therefore the exact gradient of the model's energy \emph{at
fixed features}, not the total derivative; the omitted contribution is the
feature response,
\begin{equation}
  \underbrace{\frac{dE}{d\R_A}}_{\text{total}}
  \;=\;
  \underbrace{\frac{\partial E}{\partial \R_A}
    \bigg|_{\text{features}}}_{\text{computed}}
  \;+\;
  \underbrace{\sum_{k}
    \frac{\partial E}{\partial \varphi_k}\,
    \frac{\partial \varphi_k}{\partial \R_A}}_{\text{omitted}},
  \label{eq:orbforce-split}
\end{equation}
with $\varphi_k$ the frozen features. Training against reference forces lets
the network absorb what of the omitted term the geometric pathway can express,
and the rest remains as model error, so \OrbO's forces are not guaranteed to be
curl-free. \OrbA\ removes the issue by construction, which is the main reason
the two models coexist.


\section{\OrbA: the atomic cluster expansion with conservative forces}
\label{sec:atomic}

\OrbA\ returns the graph to the nuclei and changes the electronic model
underneath it. Its descriptors come from a one-shot extended-H\"uckel
model~\cite{wh,hoffmann} evaluated in closed form \emph{inside the network's
forward pass}: the valence overlap, the Wolfsberg--Helmholz Hamiltonian, the
L\"owdin orthogonalization, the eigensolve and the descriptors are rebuilt from
the coordinates on every evaluation (lower row of Fig.~\ref{fig:gradient}).

\subsection{Design requirements}
\label{sec:atomic-motivation}

The organizing constraint is that the forces be exact gradients of the energy,
since a force field that is not the gradient of a scalar cannot be relied on
wherever an underlying potential is assumed. Equation~\eqref{eq:orbforce-split}
identifies the obstruction: any descriptor held fixed with respect to $\R$
contributes a response term that autograd cannot see, so every descriptor must
be an explicit differentiable function of $\R$ in the same computational graph
as the energy. A self-consistent featurizer is not excluded in principle: the
derivative of a converged SCF follows from differentiating through its
iterations or, at convergence, from the coupled-perturbed equations, which for a
scalar energy reduce to a single Z-vector solve as in analytic Hartree--Fock
gradients. Either route adds derivative integrals and a response solve to every
force evaluation, work of the order of the SCF itself, which a featurizer meant
to be cheap cannot afford. A non-self-consistent model gives up the accuracy of the mean field
but keeps its \emph{structure}, an overlap, a Hamiltonian, orbitals,
occupations and a density matrix, as a closed-form function of the
coordinates. Extended H\"uckel theory is the canonical such model; its absolute
energies are not used, only the qualitatively correct and smooth pattern of
bonding it produces. The basis is a minimal valence set, one $s$ function on
hydrogen and one $s$ plus three $p$ functions on every other supported element,
each a three-term contracted Gaussian from the STO-3G valence shells, and
restricting to $\ell\le1$ lets the overlap be written as a fixed sequence of
elementwise tensor operations (Sec.~\ref{si:atomic}).

\subsection{The extended-H\"uckel Hamiltonian}
\label{sec:tb-hamiltonian}

The diagonal entries are the valence-state ionization potentials of the
corresponding shell, $H_{\mu\mu} = -\mathrm{VSIP}(\mu)$, from the standard
extended-H\"uckel parameterization~\cite{hoffmann}. Those parameters, and the
constant $K$ below, were fitted with Slater-type orbitals, whereas the overlaps
here are computed over contracted STO-3G Gaussians, so the model is a close
variant of extended H\"uckel theory used as a feature generator rather than as
an electronic-structure method in its own right. The off-diagonal entries follow
the Wolfsberg--Helmholz rule~\cite{wh},
\begin{equation}
  H_{\mu\nu} = K\,S_{\mu\nu}\,\frac{H_{\mu\mu}+H_{\nu\nu}}{2},
  \qquad \mu\ne\nu,
  \qquad K = 1.75,
  \label{eq:wh}
\end{equation}
so two orbitals couple in proportion to how much they overlap and how deeply
bound they are. The whole matrix is an explicit elementwise function of
$\mathbf{S}(\R)$ and a constant table. Supported elements are H, B, C, N, O, F,
Na, Mg, Al, Si, P, S, and Cl. Since there is no self-consistent potential, the
Fock matrix and the Hamiltonian coincide,
\begin{equation}
  \mathbf{F} \equiv \mathbf{H},
  \label{eq:f-equals-h}
\end{equation}
which makes some descriptors redundant (Sec.~\ref{sec:tb-redundancy}).

\subsection{Orthogonalization, the eigenproblem, and the density matrices}
\label{sec:tb-solve}

The generalized eigenproblem $\mathbf{H}\mathbf{C} = \mathbf{S}\mathbf{C}\mathbf{e}$
is reduced to a standard one by L\"owdin orthogonalization~\cite{lowdin},
\begin{equation}
  \mathbf{X} = \mathbf{S}^{-1/2},
  \qquad
  \mathbf{H}' = \mathbf{X}\mathbf{H}\mathbf{X}
  = \mathbf{V}\,\diag(\mathbf{e})\,\mathbf{V}^\top .
  \label{eq:lowdin}
\end{equation}
The symmetric root is the orthogonalization closest to the identity, so the
orthogonalized functions stay atom-like, and it is itself a spectral function
of a symmetric matrix, so it admits the exact derivative of
Sec.~\ref{sec:spectral}. Occupations are closed-shell: with $n_{\mathrm{val}}$
valence electrons and total charge $q$,
\begin{equation}
  n_{\mathrm{occ}} = \tfrac12\big(n_{\mathrm{val}} - q\big),
  \label{eq:nocc}
\end{equation}
and $\mathrm{occ}_i = \mathbb{1}[i < n_{\mathrm{occ}}]$ over ascending
eigenvalues. An odd electron count has no valid closed-shell occupation and is
rejected rather than silently truncated. The density matrices are formed in the
orthogonal basis and transformed back,
\begin{equation}
  \mathbf{P}' = \mathbf{V}\,\diag(2\,\mathrm{occ})\,\mathbf{V}^\top,
  \qquad
  \mathbf{W}' = \mathbf{V}\,\diag(2\,\mathrm{occ}\circ\mathbf{e})\,
                \mathbf{V}^\top,
  \label{eq:tb-pw-orth}
\end{equation}
\begin{equation}
  \mathbf{P} = \mathbf{X}\mathbf{P}'\mathbf{X},
  \qquad
  \mathbf{W} = \mathbf{X}\mathbf{W}'\mathbf{X}.
  \label{eq:tb-pw}
\end{equation}
The electronic solve runs in double precision regardless of the network's
precision. Exact degeneracy itself is handled by the divided differences of
Sec.~\ref{sec:spectral}, which stay finite however close two eigenvalues are;
precision is needed to \emph{recognize} a degeneracy. In single precision,
rounding splits the exactly degenerate levels of a symmetric molecule by more
than the $10^{-8}$ threshold below which a pair is treated as degenerate, and the
divided difference becomes a ratio of rounding errors: at the symmetric
geometries of CH$_4$, NH$_3$ and C$_6$H$_6$ the single-precision descriptor
gradient is off by 5--12\%, against $10^{-5}$ after a $10^{-3}$\,{\AA}
distortion. The matrices are small enough that double precision costs almost
nothing.

\subsection{Exact differentiation of spectral functions}
\label{sec:spectral}

Three of these objects, $\mathbf{S}^{-1/2}$, $\mathbf{P}'$, and $\mathbf{W}'$,
are \emph{spectral functions}:
\begin{equation}
  \mathbf{Y} = \mathbf{V}\,\diag\!\big(f(\mathbf{w})\big)\,\mathbf{V}^\top,
  \qquad
  \mathbf{A} = \mathbf{V}\,\diag(\mathbf{w})\,\mathbf{V}^\top,
  \label{eq:specfun}
\end{equation}
for a symmetric $\mathbf{A}$ and a scalar function $f$ of its eigenvalues.
Getting their derivative wrong produces forces wrong by tens of percent on
symmetric molecules. Differentiating the eigendecomposition and assembling
$\mathbf{Y}$ fails, because first-order eigenvector perturbation theory gives
\begin{equation}
  d\mathbf{v}_i = \sum_{j\ne i}
     \frac{\mathbf{v}_j^\top d\mathbf{A}\,\mathbf{v}_i}{w_i - w_j}\,
     \mathbf{v}_j ,
  \label{eq:eigvecpert}
\end{equation}
which diverges whenever two eigenvalues coincide, the generic situation here:
the $t_2$ levels of methane, the $e$ pair of ammonia and the $\pi$ pairs of
benzene are exactly degenerate in a symmetric tight-binding Hamiltonian. The
standard workaround broadens the denominator,
\begin{equation}
  \frac{1}{d} \;\longrightarrow\; \frac{d}{d^2+\epsilon},
  \qquad d = w_i - w_j ,
  \label{eq:broadening}
\end{equation}
and for $\mathbf{W}'$ it is simply wrong: for two degenerate occupied levels it
gives a coefficient proportional to $d^2/(d^2+\epsilon)\to 0$ where the correct
limit is $2\,\mathrm{occ}$, an error that does not shrink with $\epsilon$ and
depends on which basis the backend returns inside the degenerate block
(Sec.~\ref{si:results}). The correct derivative needs no eigenvector
derivatives at all~\cite{daleckii,bhatia,higham}.

\begin{proposition}[Daleckii--Krein]
\label{prop:dk}
Let $\mathbf{A}$ be real symmetric with spectral decomposition
$\mathbf{A}=\mathbf{V}\diag(\mathbf{w})\mathbf{V}^\top$ and let $\mathbf{Y}$ be
given by Eq.~\eqref{eq:specfun}. Then the Fr\'echet derivative of the map
$\mathbf{A}\mapsto\mathbf{Y}$ is
\begin{equation}
  \begin{gathered}
  d\mathbf{Y} = \mathbf{V}\Big[
     \mathbf{L}\circ\big(\mathbf{V}^\top d\mathbf{A}\,\mathbf{V}\big)
  \Big]\mathbf{V}^\top, \\
  L_{ij} = \frac{f(w_i)-f(w_j)}{w_i-w_j},
  \quad L_{ii}=f'(w_i),
  \end{gathered}
  \label{eq:dk}
\end{equation}
with $\mathbf{L}$ the Loewner matrix of first divided differences. Because
$\mathbf{L}$ is symmetric the map is self-adjoint in the Frobenius inner
product, and the reverse-mode adjoint of a cotangent $\mathbf{g}_{\mathbf{Y}}$ is
\begin{equation}
  \begin{gathered}
  \mathbf{g}_{\mathbf{A}} =
  \mathrm{sym}\Big(
    \mathbf{V}\big[\mathbf{L}\circ
      (\mathbf{V}^\top \mathbf{g}_{\mathbf{Y}}\mathbf{V})\big]
    \mathbf{V}^\top\Big), \\
  \mathrm{sym}(\mathbf{M}) = \tfrac12(\mathbf{M}+\mathbf{M}^\top).
  \end{gathered}
  \label{eq:dk-adjoint}
\end{equation}
\end{proposition}

In Eq.~\eqref{eq:dk} the numerator of $L_{ij}$ vanishes with its denominator,
so a degeneracy is a \emph{removable} singularity with limit $f'(w)$: there is
nothing to regularize and no $\epsilon$ to choose.

\begin{proposition}[Basis independence]
\label{prop:basis}
The derivative \eqref{eq:dk} is invariant under any orthogonal rotation of
eigenvectors within a degenerate eigenspace.
\end{proposition}

\begin{proof}[Argument]
Group the indices into blocks of equal eigenvalue with spectral projectors
$\Pi_B = \sum_{i\in B}\mathbf{v}_i\mathbf{v}_i^\top$, which depend only on
$\mathbf{A}$. For $i\in B$, $j\in B'$, $L_{ij}$ depends only on the two
eigenvalues and so is a constant $L_{BB'}$, and Eq.~\eqref{eq:dk} collapses to
$d\mathbf{Y} = \sum_{B,B'} L_{BB'}\,\Pi_B\,d\mathbf{A}\,\Pi_{B'}$, expressed
entirely in projectors; the diagonal blocks enter with $f'(w_B)$.
\end{proof}

This is why the forces are reproducible across linear-algebra backends.
\OrbA\ implements
Proposition~\ref{prop:dk} as two custom differentiable operations. For the
orthogonalizer, $f(w) = w^{-1/2}$ and $f'(w) = -\tfrac12\,w^{-3/2}$, with
eigenvalues floored at $10^{-8}$, a floor a renormalized minimal valence basis
does not reach. For the density matrices, $f_P(w) = 2\,\mathrm{occ}$ and
$f_W(w) = 2\,\mathrm{occ}\,w$, with the occupation attached to the eigenvalue
\emph{index}, and their divided differences behave quite differently at a
degeneracy:
\begin{equation}
  \begin{gathered}
  L^P_{ij}
   = \frac{2\,\mathrm{occ}_i - 2\,\mathrm{occ}_j}{w_i-w_j}
   \;\xrightarrow[\ \text{degenerate}\ ]{}\; 0, \\
  L^W_{ij}
   = \frac{2\,\mathrm{occ}_i w_i - 2\,\mathrm{occ}_j w_j}{w_i-w_j}
   \;\xrightarrow[\ \text{degenerate}\ ]{}\; 2\,\mathrm{occ} .
  \end{gathered}
  \label{eq:LPLW}
\end{equation}
Degenerate levels share an occupation, so $\mathbf{P}$'s limit is zero while
$\mathbf{W}$'s is the finite $2\,\mathrm{occ}$ that broadening destroys, which is
why the failure is specific to $\mathbf{W}$ and the energy-partition features
built from it. The
linearization assumes a fixed occupation pattern, that is, no degeneracy
straddling the Fermi level, where $\mathbf{P}$ itself would be discontinuous.

Near a degeneracy the plain quotient suffers catastrophic cancellation, so the
implementation switches to the analytic limit,
\begin{equation}
  \begin{gathered}
  L_{ij} =
  \begin{cases}
    \tfrac12\big(f'(w_i)+f'(w_j)\big),
      & \begin{aligned}[t]&|w_i-w_j| < \tau\ \text{and}\\ &|f_i-f_j| < 4\tau\,m_{ij},\end{aligned}\\[8pt]
    \dfrac{f_i-f_j}{w_i-w_j}, & \text{otherwise},
  \end{cases} \\
  \tau = 10^{-8},
  \end{gathered}
  \label{eq:loewner-impl}
\end{equation}
with $m_{ij}=\max\!\big(1,|f'(w_i)|,|f'(w_j)|\big)$. The midpoint limit is
second-order accurate in the gap, and exact for the density matrices, whose
$f'$ is piecewise constant. The condition on the numerator is essential, since
the singularity is removable only when the two levels share an occupation: for
a pair straddling the Fermi level $f_i - f_j = 2$ however close the
eigenvalues, $L_{ij} = 2/(w_i-w_j)$ genuinely diverges, and a test on the gap
alone would replace it with $\tfrac12(f'+f') = 0$. Scaling the threshold with the local slope admits the removable
case for $\mathbf{S}^{-1/2}$, whose derivative can exceed $2$ in magnitude. The
same condition is why Proposition~\ref{prop:conservative} requires a frontier
gap exceeding $\tau$. In the data reported here the margin is ample: over $260$
configurations spanning all ten revised-MD17 molecules and 3BPA at $300$, $600$
and $1200$\,K the smallest frontier gap is $3.0\times 10^{-2}\,\Eh$, and the
branch is never reached across an occupied/virtual pair.

\subsection{Atom- and pair-resolved features}
\label{sec:tb-features}

With $\mathbf{P}$, $\mathbf{W}$, $\mathbf{S}$, and $\mathbf{H}$ in hand, the
descriptors are contractions of their atom blocks. With
$G_{\mu A} = \mathbb{1}[\mu \in A]$ the AO-to-atom indicator, the block sums of
any AO matrix are a single triple product,
\begin{equation}
  \big[\mathbf{G}^\top \mathbf{M}\,\mathbf{G}\big]_{AB}
  = \sum_{\mu\in A}\sum_{\nu\in B} M_{\mu\nu},
  \label{eq:blocksum}
\end{equation}
and the pairwise Frobenius norms follow from the same product on
$\mathbf{M}\circ\mathbf{M}$, so the whole featurization is a handful of dense,
differentiable matrix products. Each atom carries twelve invariant scalars
(Table~\ref{tab:tbnode}) and each atom pair six (Table~\ref{tab:tbedge});
equivariance in \OrbA\ is supplied entirely by the geometry.

\begin{table*}[!tbp]
\centering
\caption{The twelve invariant node features of \OrbA. Mulliken AO populations
are $\diag(\mathbf{P}\mathbf{S})$ and $Z^{\mathrm{val}}_A$ is the valence
electron count of the element.}
\label{tab:tbnode}
\vspace{5pt}
\small
\renewcommand{\arraystretch}{1.2}
\begin{tabular}{clll}
\toprule
index & symbol & definition & interpretation \\
\midrule
0     & $q_A$              & $Z^{\mathrm{val}}_A - p_A$ & Mulliken charge \\
1     & $p_A$              & $\sum_{\mu\in A}(\mathbf{P}\mathbf{S})_{\mu\mu}$ & total valence population \\
2--6  & $p^{(\ell)}_A$     & $\sum_{\mu\in A,\,\ell_\mu=\ell}(\mathbf{P}\mathbf{S})_{\mu\mu}$ & population per angular channel $s,p,d,f,g$ \\
7     &                    & $\Tr(\mathbf{P}_{AA}\mathbf{H}_{AA})$ & on-site one-electron energy [$\Eh$] \\
8     &                    & $\Tr(\mathbf{P}_{AA}\mathbf{F}_{AA})$ & on-site Fock energy [$\Eh$] \\
9     &                    & $\Tr(\mathbf{W}_{AA})$ & on-site energy-weighted diagonal [$\Eh$] \\
10    &                    & $\|\mathbf{P}_{AA}\|_F$ & local density magnitude \\
11    &                    & valence population & population excluding core \\
\bottomrule
\end{tabular}
\end{table*}

\begin{table*}[!tbp]
\centering
\caption{The six invariant edge features of \OrbA, for an ordered atom pair
$(A,B)$.}
\label{tab:tbedge}
\vspace{5pt}
\small
\renewcommand{\arraystretch}{1.2}
\begin{tabular}{lll}
\toprule
index & definition & interpretation \\
\midrule
0 & $\sum_{\mu\in A,\,\nu\in B}(\mathbf{P}\mathbf{S})_{\mu\nu}(\mathbf{P}\mathbf{S})_{\nu\mu}$ & Mayer bond order \\
1 & $\|\mathbf{P}_{AB}\|_F$ & interatomic density coupling \\
2 & $\|\mathbf{F}_{AB}\|_F$ & Fock coupling magnitude \\
3 & $\|\mathbf{S}_{AB}\|_F$ & overlap magnitude \\
4 & $\|\mathbf{H}_{AB}\|_F$ & core-Hamiltonian coupling \\
5 & $\Tr(\mathbf{P}_{AB}\mathbf{F}_{BA})$ & pair bond-energy coupling \\
\bottomrule
\end{tabular}
\end{table*}

\begin{figure*}[!tbp]
\centering
\includegraphics[width=\linewidth]{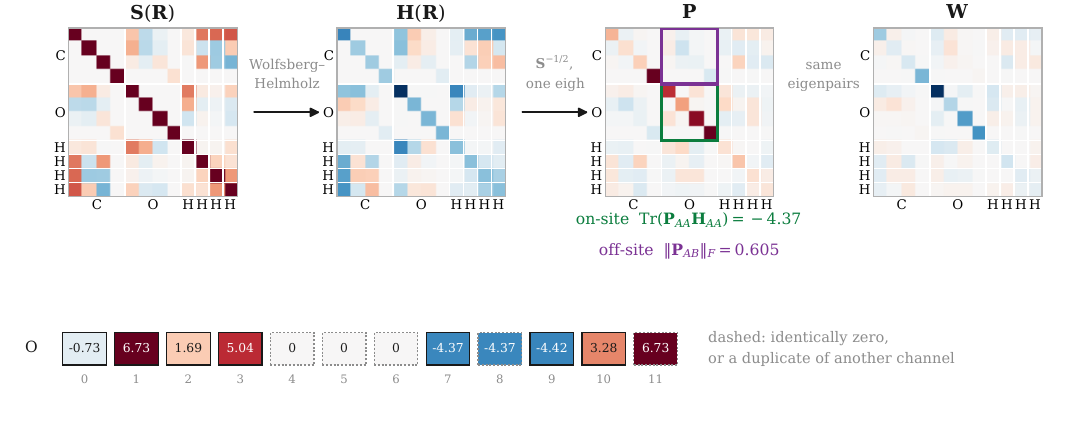}
\caption{How \OrbA's descriptors are built, on methanol in its minimal valence
basis ($n_{\mathrm{AO}}=12$). The four matrices are the featurizer's own
output: the overlap $\mathbf{S}(\mathbf{R})$ of
Eq.~\eqref{eq:contracted-ovlp}, the Wolfsberg--Helmholz Hamiltonian of
Eq.~\eqref{eq:wh}, and the density and energy-weighted density matrices of
Eq.~\eqref{eq:tb-pw}, both from the single eigendecomposition of
Eq.~\eqref{eq:lowdin}. White rules mark the atom blocks, and each panel is
colour-scaled independently. Every descriptor is a trace or Frobenius norm of
one block, on-site blocks giving node descriptors and off-site blocks edge
descriptors, which is the source of the rotational invariance of
Proposition~\ref{prop:tb-invariant}. The strip gives the twelve node channels on
oxygen: channels 4--6 are zero for want of $d$, $f$, $g$ functions, and
channels 8 and 11 duplicate 7 and 1, leaving five independent scalars.}
\label{fig:featurization}
\end{figure*}

The Mayer bond order~\cite{mayer} of edge feature 0 recovers approximately $1$,
$2$, and $3$ for single, double, and triple bonds and decays smoothly with
distance, a far more chemically direct edge descriptor than a distance. Node
features 7--9 are energy-like: feature 7, identical to 8 since
$\mathbf{F}\equiv\mathbf{H}$, partitions $\Tr(\mathbf{P}\mathbf{H})$ together
with the off-site coupling of edge feature 5 wherever an edge exists, so the
scale of these channels carries physical information, which is why
Sec.~\ref{sec:atom-embed} standardizes against dataset statistics rather than
per sample. Edge feature 5 is signed, so it distinguishes bonding from
antibonding character. Because $\mathbf{F}\equiv\mathbf{H}$ and the basis is
valence-only with no $d$, $f$, or $g$ functions, the independent electronic
content given the element is five scalars per atom and five per edge
(Sec.~\ref{sec:tb-redundancy}); molecules are featurized in batches by
block-diagonal padding with results identical to the per-molecule path
(Sec.~\ref{sec:tb-batching}). Figure~\ref{fig:featurization} shows the
construction on methanol.

\subsection{Readout, energy, and baseline}
\label{sec:atom-energy}

Because nodes are atoms, no pooling is needed, and the energy decomposition is
\begin{equation}
  E = \underbrace{\sum_A E^{0,\mathrm{atom}}_{Z_A}}_{E_0}
    + \underbrace{\sum_A E^{\mathrm{int}}_A}_{E_{\mathrm{inter}}}
    + E_{\mathrm{ZBL}}
    + E_{\mathrm{LR}} ,
  \label{eq:atom-energy}
\end{equation}
with a per-element baseline only and no correction head. Every term other than
$E_0$ is geometry-dependent and enters the force.

\subsection{Conservative forces}
\label{sec:conservative}

\begin{proposition}
\label{prop:conservative}
Let $E(\R)$ be the energy of Eq.~\eqref{eq:atom-energy} with descriptors
computed by Secs.~\ref{sec:tb-overlap}--\ref{sec:tb-features}. On any open set
of configurations where the overlap matrix is positive definite and the gap
between the highest occupied and lowest unoccupied tight-binding level exceeds
the divided-difference tolerance $\tau$ of Eq.~\eqref{eq:loewner-impl},
$E$ is a differentiable function of $\R$, and the force returned by
reverse-mode differentiation satisfies
$\mathbf{F}_A = -\partial E/\partial\R_A$ exactly. Consequently
$\nabla\times\mathbf{F} = 0$ and the dynamics generated by $(E,\mathbf{F})$
conserves $E + T$.
\end{proposition}

\begin{proof}[Argument]
$E$ is a composition of maps each differentiable on the stated set: the
contracted overlap \eqref{eq:contracted-ovlp} is a polynomial-times-Gaussian
function of the coordinates; the Hamiltonian \eqref{eq:wh} is elementwise
algebraic in it; $\mathbf{S}^{-1/2}$ and the density matrices are spectral
functions, differentiable by Proposition~\ref{prop:dk} wherever the occupation
pattern is fixed, which the gap assumption guarantees; the features
\eqref{eq:blocksum} are polynomial contractions; and the network, the ZBL term,
the readout, and the monopole electrostatic term when enabled are
differentiable by construction. No quantity in the chain is detached from the
coordinates, so reverse-mode differentiation returns the exact gradient and
hence a curl-free field admitting $E$ as potential.
\end{proof}

Numerically, the analytic force agrees with a central difference of the model's
own energy to better than $10^{-4}$\,eV/{\AA}, the discretization floor of the
check. The two caveats are physical: a singular overlap would mean a linearly
dependent basis, which a renormalized minimal valence basis does not produce,
and a level crossing at the Fermi level makes the density matrix itself
discontinuous, so no potential built on it can be smooth there.

\subsection{Domain of validity}
\label{sec:tb-limits}

The featurizer is deliberately narrow: closed-shell only, which excludes
radicals, triplets and most transition-metal centers, and main-group $s/p$
only, since the parameter tables and the overlap of Eq.~\eqref{eq:s1d} assume
$\ell\le 1$. Being non-self-consistent it performs no charge equilibration and
describes long-range charge transfer and strongly ionic bonding poorly, although
the network can compensate where the deficiency is a smooth function of what it
sees. Within this domain the descriptors are smooth, gauge-independent, cheap,
and exactly differentiable. Every reported benchmark lies within a narrower domain still: neutral,
closed-shell, non-periodic molecules of H, C, N and O.


\section{Equivariance of the featurized models}
\label{sec:equivariance}

The backbone is equivariant by construction, but a descriptor that is not a
proper irrep, placed in a channel that assumes one, breaks the symmetry of the
whole model, and a dataset without rotated copies of its structures will not
reveal it. This section shows that every descriptor is a genuine invariant or a
proper irrep.

\subsection{How the atomic-orbital basis transforms}
\label{sec:ao-transform}

Under $\R_A \mapsto \mathbf{Q}\R_A$ with $\mathbf{Q}\in O(3)$, atom-centred
shells of definite angular momentum carry the basis into itself,
\begin{equation}
  \chi_\mu \;\longmapsto\; \sum_\nu D_{\mu\nu}(\mathbf{Q})\,\chi_\nu ,
  \label{eq:ao-rot}
\end{equation}
with $\mathbf{D}(\mathbf{Q})$ orthogonal and block diagonal in the atom index,
$[\mathbf{D}]_{AB}=\delta_{AB}\mathbf{D}_A$. Orthogonality holds in both
models: \OrbO\ uses real solid-harmonic orbitals, whose degree-$\ell$ block is
the real Wigner matrix, and \OrbA\ a Cartesian $s/p$ basis, whose $p$ block is
$\mathbf{Q}$ itself and whose contracted renormalization scales the three $p$
functions of a shell equally. (Cartesian $d$ functions would not be
orthogonal, one reason the tight-binding basis stops at $\ell=1$.) Every AO
matrix built from the geometry then transforms by congruence,
\begin{equation}
  \begin{gathered}
  \mathbf{S} \mapsto \mathbf{D}\mathbf{S}\mathbf{D}^\top, \\
  \mathbf{H} \mapsto \mathbf{D}\mathbf{H}\mathbf{D}^\top, \\
  \mathbf{F} \mapsto \mathbf{D}\mathbf{F}\mathbf{D}^\top, \\
  \mathbf{P} \mapsto \mathbf{D}\mathbf{P}\mathbf{D}^\top, \\
  \mathbf{W} \mapsto \mathbf{D}\mathbf{W}\mathbf{D}^\top ,
  \end{gathered}
  \label{eq:congruence}
\end{equation}
and each atom-pair block using only its own two blocks,
\begin{equation}
  \mathbf{M}_{AB} \;\longmapsto\;
  \mathbf{D}_A\,\mathbf{M}_{AB}\,\mathbf{D}_B^\top .
  \label{eq:block-rot}
\end{equation}
Overlap integrals depend only on coordinate differences, so every descriptor
below is translation invariant as well; \OrbO's second moment
$\langle r^2\rangle_i$, whose operator carries an origin, is treated in
Sec.~\ref{si:equiv}.

\subsection{The atomic model's descriptors are invariants}
\label{sec:equiv-atomic}

\begin{proposition}
\label{prop:tb-invariant}
Every one of the twelve node descriptors of Table~\ref{tab:tbnode} and the six
edge descriptors of Table~\ref{tab:tbedge} is invariant under the full
orthogonal group $O(3)$ acting on the nuclear coordinates, and under
translations.
\end{proposition}

\begin{proof}[Argument]
By cyclicity and $\mathbf{D}^\top\mathbf{D}=\mathbf{I}$,
\begin{equation}
  \begin{gathered}
  \Tr\!\big(\mathbf{D}_A\mathbf{X}\mathbf{D}_A^\top\big) = \Tr\mathbf{X}, \\
  \Tr\!\big(\mathbf{D}_A\mathbf{X}\mathbf{D}_B^\top\,
            \mathbf{D}_B\mathbf{Y}\mathbf{D}_A^\top\big)
    = \Tr(\mathbf{X}\mathbf{Y}),
  \end{gathered}
  \label{eq:block-rot-traces}
\end{equation}
and $\|\mathbf{D}_A\mathbf{X}\mathbf{D}_B^\top\|_F = \|\mathbf{X}\|_F$. The
populations are traces of $(\mathbf{P}\mathbf{S})_{AA}$, restricted to a shell
block for the per-$\ell$ channels, which is legitimate because $\mathbf{D}_A$ is
block diagonal by shell; the on-site energies are traces of on-site blocks; the
Mayer bond order and the pair coupling are two-block traces; and the remaining
descriptors are block norms.
\end{proof}

Every expression is quadratic in $\mathbf{D}$, so improper rotations are covered
too, and the descriptors are indexed by atom and atom pair, so permutation
equivariance holds; since every descriptor lands in a $0e$ channel, \OrbA's
equivariance is exactly that of the bare backbone. The argument for \OrbO,
whose molecular-orbital matrix elements are invariant because the two factors
of $\mathbf{D}$ cancel as in Eq.~\eqref{eq:congruence} and whose centroid,
polarization and traceless quadrupole are $1o$ and $2e$ objects, is given in
Sec.~\ref{si:equiv}.


\section{Delta learning and multi-fidelity decomposition}
\label{sec:delta}

\begin{figure*}[!tbp]
\centering
\includegraphics[width=\linewidth]{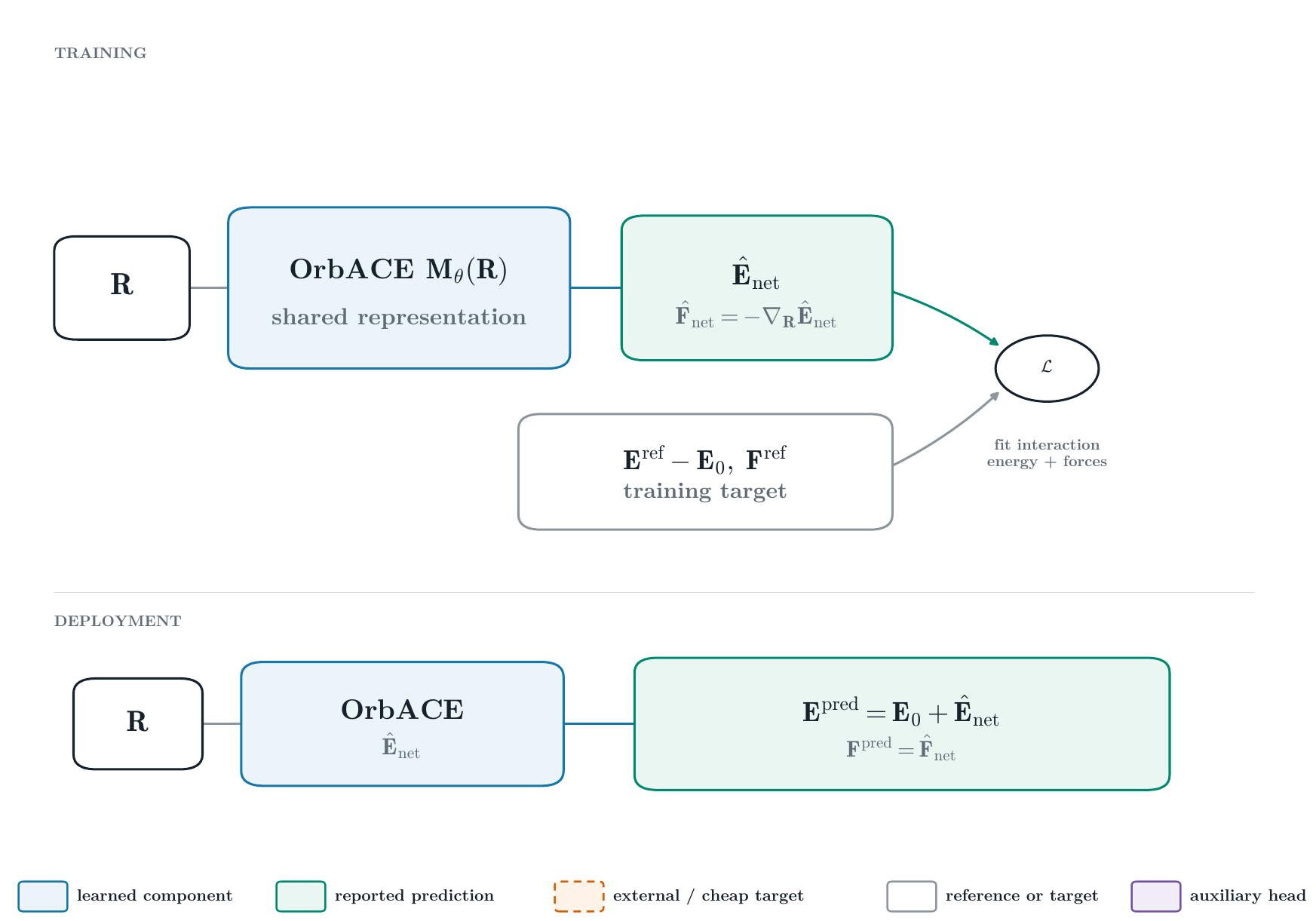}
\caption{The \texttt{direct} scheme. No cheap level appears at any stage; the
network is fitted to the reference with the isolated-atom term removed, and a
prediction is one network pass.}
\label{fig:scheme-direct}
\end{figure*}

\begin{figure*}[!tbp]
\centering
\includegraphics[width=\linewidth]{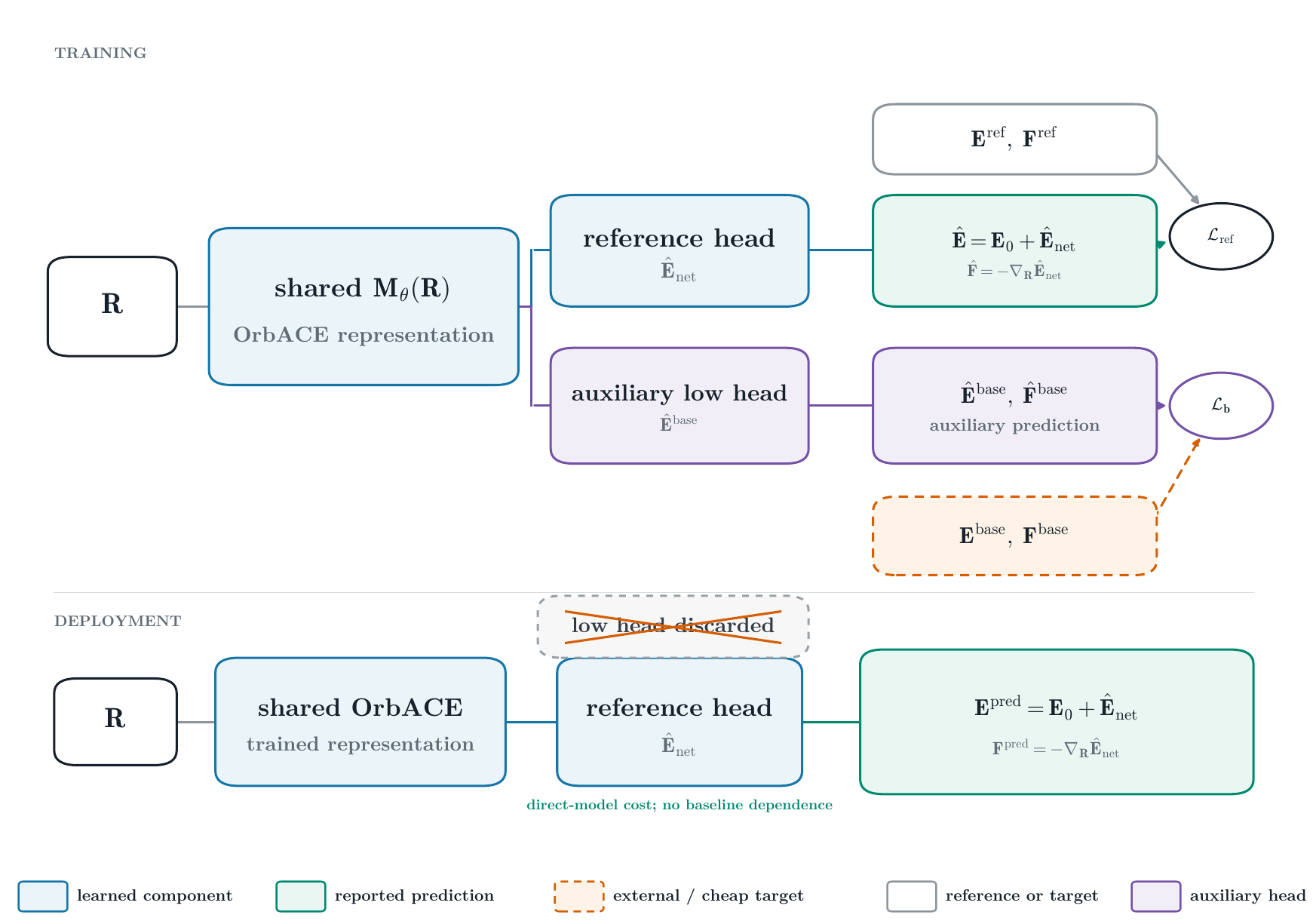}
\caption{The parallel multi-fidelity scheme, \texttt{mfah}. The auxiliary head
never enters the total, so the high-fidelity path is that of
Fig.~\ref{fig:scheme-direct} and the head is deleted once training ends; the
cheap level reaches the prediction only through the shared representation.}
\label{fig:scheme-mfah}
\end{figure*}

Both models can be trained on absolute targets or on residuals relative to a
cheap electronic-structure baseline. The organizing question is what plays the
role of the \emph{low-level function} $E^{\mathrm{low}}$: whether a cheap level
exists at all, whether it is an external code or a head of the network, and
whether it contributes to the predicted energy. With $E_0 = \sum_A E_0(Z_A)$
the frozen isolated-atom reference and $E^{\mathrm{net}}$ the network's
interaction energy, the four training schemes are set out in
Table~\ref{tab:low-schemes} and drawn in Figs.~\ref{fig:scheme-direct}--%
\ref{fig:scheme-mfah}, each showing training above and deployment below.

\begin{table*}[!tbp]
\centering
\caption{The four training schemes, distinguished by their low-level function
$E^{\mathrm{low}}$. The force target follows the energy target in every row.
``Fitted to'' is the target for the network's own output; the two
multi-fidelity rows additionally anchor $E^{\mathrm{low}}$ to the stored cheap
level $E^{\mathrm{base}}$. The last column is what the cheap level costs at
deployment.}
\label{tab:low-schemes}
\vspace{5pt}
\small
\begin{tabular}{@{}lllll@{}}
\toprule
Scheme & $E^{\mathrm{low}}$ & $E^{\mathrm{pred}}$ & Fitted to
  & Cheap level at inference \\
\midrule
\texttt{direct}  & none
  & $E_0 + E^{\mathrm{net}}$
  & $E^{\mathrm{ref}} - E_0$ & absent \\
\texttt{regular} & $E^{\mathrm{base}}$, external
  & $E_0 + E^{\mathrm{low}} + E^{\mathrm{net}}$
  & $E^{\mathrm{ref}} - E_0 - E^{\mathrm{base}}$ & recomputed \\
\texttt{mfdh}    & $E^{\mathrm{base}}_{\mathrm{net}}$, learned
  & $E_0 + E^{\mathrm{low}} + E^{\mathrm{delta}}_{\mathrm{net}}$
  & $E^{\mathrm{ref}} - E_0$ & evaluated \\
\texttt{mfah}    & $E^{\mathrm{low}}_{\mathrm{net}}$, learned
  & $E_0 + E^{\mathrm{net}}$
  & $E^{\mathrm{ref}} - E_0$ & discarded \\
\bottomrule
\end{tabular}
\end{table*}

\begin{figure*}[!tbp]
\centering
\includegraphics[width=\linewidth]{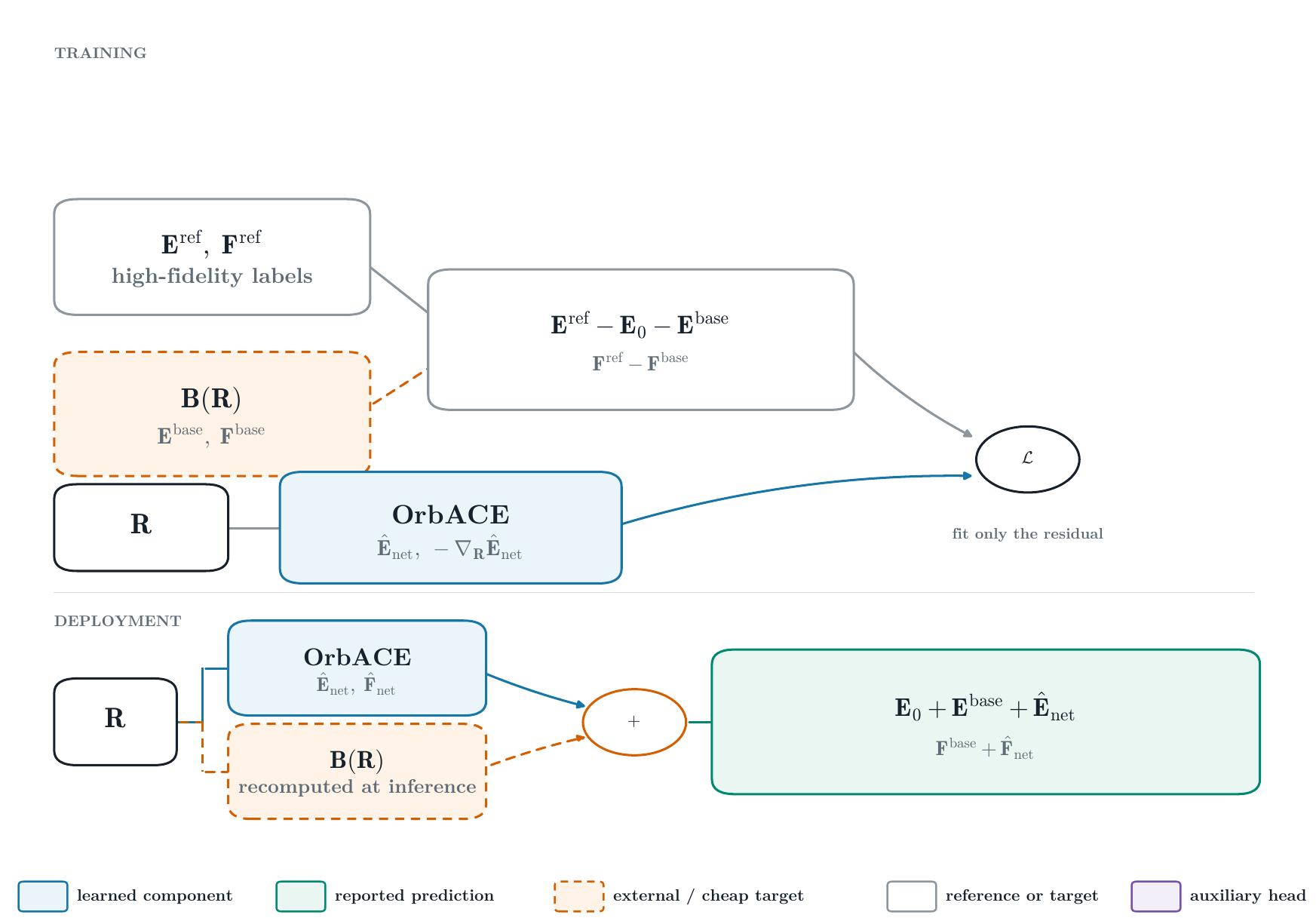}
\caption{The \texttt{regular} residual scheme. The external baseline $B(\R)$ is used to construct the training targets and must also be evaluated at every inference call, adding its computational cost to that of the network. Each loss node stands for an energy and a force
term in this and the following figures.}
\label{fig:scheme-regular}
\end{figure*}

Only two objectives are involved (Sec.~\ref{sec:training}): \texttt{direct} and
\texttt{regular} share one loss and differ only in the target, and
\texttt{mfdh} and \texttt{mfah} share a four-term loss and differ only in
whether $E^{\mathrm{low}}$ is an addend of $E^{\mathrm{pred}}$. In
\texttt{mfdh}, the multi-fidelity \emph{delta head}, the high-fidelity output
is a delta added to a learned base; in \texttt{mfah}, the multi-fidelity
\emph{absolute head}, an extra head predicts the cheap level's absolute energy,
shapes the shared representation, and is discarded.

\subsection{Targets and the direct case}

After removal of the isolated-atom baseline the stored labels are
\begin{equation}
  \begin{gathered}
  E^{\mathrm{store}} = E^{\mathrm{ref}}
    - \sum_A E_0(Z_A)
    - E^{\mathrm{base}}, \\
  \mathbf{F}^{\mathrm{store}} = \mathbf{F}^{\mathrm{ref}}
    - \mathbf{F}^{\mathrm{base}},
  \end{gathered}
  \label{eq:delta-targets}
\end{equation}
with $E^{\mathrm{base}} = 0$ and $\mathbf{F}^{\mathrm{base}} = \mathbf{0}$ in
the \emph{direct} case. Each prepared frame records its baseline energy and forces, and the dataset
which baseline was used. Available baselines span the GFN$n$-xTB
methods~\cite{gfn2}, Hartree--Fock and Kohn--Sham density-functional theory;
residual training, $\Delta$-machine learning~\cite{deltaml}, works because a
baseline that captures the bonding topology leaves the network a smaller and
smoother correction to fit.

\subsection{The \texttt{regular} scheme and its cost at inference}

In the \texttt{regular} scheme the network predicts the residual, and at
inference the baseline is recomputed and added back:
\begin{equation}
  \begin{gathered}
  E^{\mathrm{pred}} = \sum_A E_0(Z_A)
    + E^{\mathrm{base}}(\R)
    + E^{\mathrm{net}}(\R), \\
  \mathbf{F}^{\mathrm{pred}} = \mathbf{F}^{\mathrm{base}}(\R)
    + \mathbf{F}^{\mathrm{net}}(\R).
  \end{gathered}
  \label{eq:regular}
\end{equation}
The baseline must be evaluated at every energy call. For \OrbO\ a Hartree--Fock
baseline at the feature basis comes from the same mean field that produces the
features; for \OrbA\ it reintroduces per step the self-consistent calculation
its featurizer exists to avoid, and an xTB baseline is cheap but still an
external call.

\OrbA\ does receive a cheap-level energy for nothing: node feature 7 of
Table~\ref{tab:tbnode} and edge feature 5 of Table~\ref{tab:tbedge} together
partition $\Tr(\mathbf{P}\mathbf{H})$, the extended-H\"uckel band energy, per
atom and per bond. But a non-self-consistent H\"uckel energy, with no charge
equilibration and no repulsion, is not a \emph{good} cheap level and is too
inaccurate to serve as the subtrahend of a residual; a learned base head
upgrades it to a level that would otherwise require a self-consistent
calculation.

\begin{remark}
\label{rem:hfgrad}
With a Hartree--Fock baseline, \OrbO\ must evaluate the baseline's analytic
nuclear gradient on the \emph{canonical} mean field, before localization, since
localization changes the energy-weighted density matrix that enters the
gradient. The baseline \emph{energy} is a byproduct of the converged mean
field; its gradient is not, as it requires derivative integrals, so a scheme
that consumes baseline forces pays for them even when the energy is free.
\end{remark}

\subsection{The multi-fidelity schemes}
\label{sec:mf}

Both multi-fidelity schemes replace the external low-level function with a head
of the network, trained to reproduce the cheap level from the same features the
high-fidelity path uses, so nothing outside the network is evaluated at
inference. We call it the \emph{cheap head} in both modes and the \emph{base
head} in \texttt{mfdh}, where it is an addend of the reported energy. The heads
read the final scalar node features; in \OrbA\ they also read the raw
standardized tight-binding scalars through a skip connection, and in \OrbO\ the
cheap head is a per-atom perceptron on the Mulliken-pooled features of
Eq.~\eqref{eq:atompool}.

A learned cheap head removes a per-configuration call to an external code at
the price of capacity spent on a quantity the reference does not supervise, so
it is worth having only when the cheap level is a \emph{separate} calculation
from the featurizer's. Three of the four combinations used here satisfy this:
\OrbA\ runs no self-consistent calculation, and \OrbO's xTB baseline is a
different method from its features. \OrbO's Hartree--Fock baseline, read off
the mean field it already has, does not, and Sec.~\ref{sec:ani-baseline} shows
the consequence.

\subsection{\texttt{mfdh}: the low-level function inside the total}
\label{sec:mfdh}

In the additive mode the energy is decomposed into a learned base head, softly
anchored to the cheap fidelity, and a delta head:
\begin{equation}
  E^{\mathrm{pred}} = \underbrace{E_0}_{\text{frozen}}
    + \underbrace{E^{\mathrm{base}}_{\mathrm{net}}}_{\text{learned surrogate}}
    + \underbrace{E^{\mathrm{delta}}_{\mathrm{net}}}_{\text{correction}} .
  \label{eq:mf-energy}
\end{equation}
The base head is load-bearing: it is an addend of the quantity supervised
against the reference, so the high-fidelity term and the cheap-level anchor
contend over one prediction, and the head is evaluated at every energy call.
The head that additively refines the high-fidelity readout, the delta head in
\OrbA\ and the base head in \OrbO, has its output layer zeroed, so training
starts from the non-multi-fidelity prediction.

\begin{figure*}[!tbp]
\centering
\includegraphics[width=\linewidth]{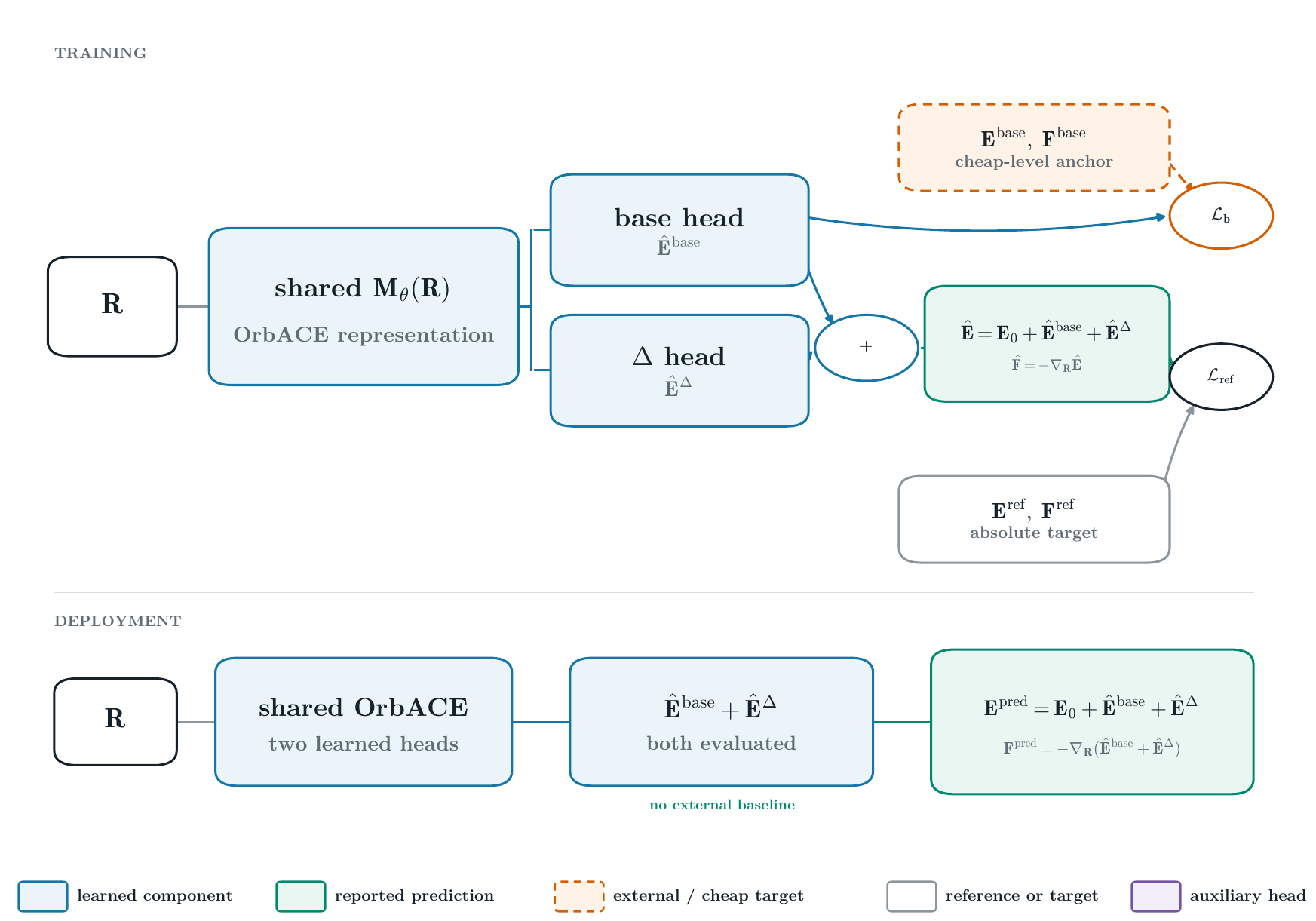}
\caption{The additive multi-fidelity scheme, \texttt{mfdh}. The learned base
head is summed into the reported energy, so it must be evaluated at
deployment, and the two energy terms of Eq.~\eqref{eq:mf-loss} act on
quantities sharing that head.}
\label{fig:scheme-mfdh}
\end{figure*}

\subsection{\texttt{mfah}: the low-level function as an auxiliary task}
\label{sec:mfah}

In the parallel mode the cheap head predicts the cheap level's \emph{absolute}
energy and never enters the total. With $\varepsilon^{\mathrm{low}}_A$ the
head's per-atom output,
\begin{equation}
  \begin{gathered}
  E^{\mathrm{pred}} = E_0 + E^{\mathrm{net}}, \\
  E^{\mathrm{low}} = E^{\mathrm{low}}_0 + \sum_A \varepsilon^{\mathrm{low}}_A, \\
  \mathbf{F}^{\mathrm{low}}
    = -\nabla_{\R} \sum_A \varepsilon^{\mathrm{low}}_A ,
  \end{gathered}
  \label{eq:mfah}
\end{equation}
where $E^{\mathrm{low}}_0$ is a frozen per-element offset carried only in the
anchor term. The two predictions share only the backbone. Deleting the head
after training therefore recovers a direct model exactly, with no baseline
dependence at deployment; the cheap level reaches the prediction only through
the shared trunk, encouraging the decomposition that \texttt{regular} and
\texttt{mfdh} impose arithmetically; the two energy terms act on disjoint
outputs, so the anchor weight no longer trades against high-fidelity accuracy;
and an \texttt{mfah} model differs from a direct one only in the attached head,
so their difference measures what the auxiliary task contributes. The price is
that \texttt{mfah} gives up \texttt{mfdh}'s interpretable decomposition and
guarantees no use of the cheap level at all. For \OrbA\ both modes are
conservative in the sense of Proposition~\ref{prop:conservative}; for \OrbO\
both inherit the fixed-feature caveat of Sec.~\ref{sec:orb-forces}.

\section{Training}
\label{sec:training}

\subsection{Objective}
\label{sec:objective}

Outside the multi-fidelity schemes the objective is a weighted sum of a
per-atom energy term and a normalized force term, optionally with stress:
\begin{equation}
  \begin{aligned}
  \mathcal{L} ={}& \lambda_E \left\langle
      \left(\frac{\hat E - E}{N_{\mathrm{at}}}\right)^{\!2}\right\rangle\\
    &+ \lambda_F \left\langle
      \left(\frac{\hat{\mathbf{F}} - \mathbf{F}}{\varsigma}\right)^{\!2}
      \right\rangle
    + \lambda_\sigma \big\langle
      (\hat{\bm\sigma}-\bm\sigma)^2\big\rangle ;
  \end{aligned}
  \label{eq:loss}
\end{equation}
the stress term is included for completeness; we set $\lambda_\sigma=0$ in all experiments reported here, which involve isolated molecules.
Dividing the energy error by the atom count keeps large molecules from
dominating, and dividing the force error by the training-set root-mean-square
force $\varsigma$ (Sec.~\ref{sec:stats}) makes it dimensionless, so the weights
express a relative priority rather than a unit conversion. A Huber
variant~\cite{huber} with threshold $\delta$, scaled by $\varsigma$ for forces,
is the default for \OrbA. Default weights are $\lambda_E=1$, $\lambda_F=10$ for
\OrbO\ and $\lambda_E=10$, $\lambda_F=100$ for \OrbA, with the ANI-1x weights in
Table~\ref{tab:runconfig}; forces dominate because they give $3N$ constraints
per structure against the energy's one.

\subsection{The multi-fidelity objective}
\label{sec:mf-loss}

Both multi-fidelity modes are trained with one four-term objective; only the
model differs:
\begin{equation}
  \begin{aligned}
  \mathcal{L}^{\mathrm{MF}} ={}&
      \lambda_E\,\ell\!\left(\frac{\hat E}{N_{\mathrm{at}}},
                             \frac{E^{\mathrm{ref}}}{N_{\mathrm{at}}}\right)\\
    &+ \lambda_F\,\ell\!\left(\frac{\hat{\mathbf{F}}}{\varsigma},
                             \frac{\mathbf{F}^{\mathrm{ref}}}{\varsigma}\right)\\
    &+ \lambda_E^{\mathrm{b}}\,
      \ell\!\left(\frac{\hat E^{\mathrm{base}}}{N_{\mathrm{at}}},
                  \frac{E^{\mathrm{base}}}{N_{\mathrm{at}}}\right)\\
    &+ \lambda_F^{\mathrm{b}}\,
      \ell\!\left(\frac{\hat{\mathbf{F}}^{\mathrm{base}}}{\varsigma},
                  \frac{\mathbf{F}^{\mathrm{base}}}{\varsigma}\right),
  \end{aligned}
  \label{eq:mf-loss}
\end{equation}
with $\ell$ a Huber loss by default and defaults $\lambda_E = 10$,
$\lambda_F = 100$, $\lambda_E^{\mathrm{b}} = 10$, $\lambda_F^{\mathrm{b}} = 1$.
The hatted base quantities are the learned low-level function
$E^{\mathrm{low}}$ of Table~\ref{tab:low-schemes} and its gradient; under
\texttt{mfdh} the same head also contributes to $\hat E$ in the first two
terms, and under \texttt{mfah} it does not.

The first two terms compare the \emph{total} prediction against the
\emph{absolute} reference, so the network never fits residual forces, a
difference of two comparable fields that is much noisier in relative terms than
either. For the same reason $\lambda_F^{\mathrm{b}} \ll \lambda_F$: anchoring
tightly to the baseline's forces would drag the representation toward the
baseline's errors. Because frames prepared with a baseline store residual
labels (Eq.~\ref{eq:delta-targets}), the dataset adds the baseline back in
multi-fidelity mode, and an option drops the second term for energy-only
references.

\subsection{Partially labelled data}
\label{sec:mf-partial}

The two pairs of terms read different labels, and the cheap level is
affordable on far more configurations than the expensive one. Let the training
set contain $N$ frames carrying cheap labels, of which a subset $\mathcal{R}$ of
$M \le N$ also carries reference labels, and write $\rho = M/N$. For a batch
$\mathcal{B}$ the objective is
\begin{equation}
  \begin{aligned}
  \mathcal{L}^{\mathrm{MF}}(\mathcal{B}) ={}&
    \underbrace{\lambda_E\,\big\langle \ell^{E}_i \big\rangle_{\mathcal{B}\cap\mathcal{R}}
              + \lambda_F\,\big\langle \ell^{F}_i \big\rangle_{\mathcal{B}\cap\mathcal{R}}}
              _{\text{reference terms}}
    \\ &+
    \underbrace{\lambda^{\mathrm{b}}_E\,\big\langle \ell^{E,\mathrm{b}}_i \big\rangle_{\mathcal{B}}
              + \lambda^{\mathrm{b}}_F\,\big\langle \ell^{F,\mathrm{b}}_i \big\rangle_{\mathcal{B}}}
              _{\text{anchor terms}} ,
  \end{aligned}
  \label{eq:mf-partial}
\end{equation}
with the four per-frame terms of Eq.~\eqref{eq:mf-loss} and
$\langle\cdot\rangle_{\mathcal{S}}$ a mean over $\mathcal{S}$: the anchor terms
run over the whole batch and the reference terms over its labelled part.
Missing reference labels are zero-filled with a mask $m_i$, per frame for
energies and per atom for forces, and each reference mean is
\begin{equation}
  \big\langle x_i \big\rangle_{\mathcal{B}\cap\mathcal{R}}
    = \frac{\sum_{i\in\mathcal{B}} m_i\,x_i}{\sum_{i\in\mathcal{B}} m_i} .
  \label{eq:masked-mean}
\end{equation}
Dividing by $\sum_i m_i$ rather than $|\mathcal{B}|$ makes the high-fidelity
gradient independent of $\rho$, where batch-size normalization would count the
zero-filled entries as perfect predictions; a fully labelled dataset recovers
Eq.~\eqref{eq:mf-loss}, and a batch with no labelled frame contributes an exact
zero attached to the autograd graph. Under plain shuffling
$|\mathcal{B}\cap\mathcal{R}|$ is binomial with mean $\rho|\mathcal{B}|$, so at
small $\rho$ most batches carry no reference; a label-balanced sampler fixes
$|\mathcal{B}\cap\mathcal{R}| = k$ by drawing from separate streams, and is
enabled automatically when $\rho|\mathcal{B}|$ falls below $k=8$.

The multi-fidelity path computes two gradients per step and so retains the
feature graph between two backward passes, a training cost \texttt{mfah} incurs
as well. The rest of the protocol is given in Sec.~\ref{si:training}, the data
and inference pipelines in Sec.~\ref{sec:pipeline}, and the cost analysis in
Sec.~\ref{sec:complexity}.


\section{Evaluation protocol and results}
\label{sec:results}

Three benchmarks are reported. ANI-1x
(Secs.~\ref{sec:ani-data}--\ref{sec:mf-scale}) tests breadth of chemistry and
carries the primary comparison of the two models, revised MD17
(Sec.~\ref{sec:rmd17}) the small-data, single-molecule regime where an
electronic prior should matter most, and 3BPA (Sec.~\ref{sec:3bpa}) transfer
to configurations far from the training distribution.

\subsection{Benchmark and data selection}
\label{sec:ani-data}

The first benchmark is ANI-1x~\cite{ani1x} at
$\omega$B97X~\cite{wb97x}/6-31G$^*$~\cite{pople631g}, covering H, C, N, and O,
a reference unrelated to the basis of either featurizer. Splits were carved by
farthest-point sampling in the geometry descriptor space of Eq.~\eqref{eq:fps}
(Table~\ref{tab:splits}), which is what makes training on $10{,}000$
configurations and testing on $442{,}000$ meaningful: a random training set
from a trajectory-derived pool would be redundant and leave near-duplicates in
the test set.

\begin{table*}[!tbp]
\centering
\caption{ANI-1x $\omega$B97X/6-31G$^*$ splits. Each split also carries the
four \texttt{IsolatedAtom} frames (H, C, N, O) that fix the $E_0$ convention,
excluded from all reported metrics.}
\label{tab:splits}
\vspace{5pt}
\small
\begin{tabular}{lrl}
\toprule
Split & Frames & Role \\
\midrule
train    & 10,000  & model fitting \\
valid    & 1,000   & early stopping, best-checkpoint selection \\
test     & 442,000 & held out; all reported metrics \\
\midrule
\multicolumn{3}{l}{Composition: H, C, N, O. Atoms per frame: mean 13.81, range 2--55.} \\
\bottomrule
\end{tabular}
\end{table*}

Every model takes its isolated-atom energies from the dataset's own
\texttt{IsolatedAtom} frames. GFN2-xTB failed to converge on 5 training and 94
test frames, so every xTB-delta arm is trained on $9{,}995$ and evaluated on
$441{,}906$ frames, the same frames for all three models.

\subsection{Models compared}
\label{sec:ani-models}

Three models are compared on byte-identical splits, with the configurations of
Table~\ref{tab:runconfig}. The reference is \textbf{MACE}~\cite{mace},
configured to match \OrbA\ in backbone, hidden irreps, depth, correlation
order, cutoff, radial basis, loss and loss-weight schedule, optimizer, learning
rate, batch size, seed and training data, but receiving only atomic numbers and
coordinates.

\begin{table*}[!tbp]
\centering
\caption{Run configurations on ANI-1x. All three models use correlation order
$\nu=3$ and are trained for up to 1000 epochs (MACE: early stopping at 623 to
655 epochs).}
\label{tab:runconfig}
\vspace{5pt}
\footnotesize
\renewcommand{\arraystretch}{1.2}
\begin{tabular}{llll}
\toprule
 & \textbf{MACE} & \textbf{\OrbO} & \textbf{\OrbA} \\
\midrule
Node features      & atomic number only & HF orbital features & extended-H\"uckel features \\
Graph nodes        & atoms & localized MOs & atoms \\
Feature basis      & n/a & STO-3G, Pipek--Mezey & minimal valence \\
Hidden irreps      & $64\times0e{+}64\times1o{+}64\times2e$ & $64\times0e{+}64\times1o{+}64\times2e$ & $64\times0e{+}64\times1o{+}64\times2e$ \\
$\ell_{\max}$ (node) & 2 & 2 & 2 \\
Harmonics on edges & $\ell\le2$ & $\ell\le2$ & $\ell\le2$ \\
Layers $L$         & 3 & 2 & 3 \\
Cutoff $r_{\max}$  & 5.0\,{\AA} & 6.0\,{\AA} & 5.0\,{\AA} \\
Loss               & Huber, $\delta=0.01$ & MSE & Huber, $\delta=0.01$ \\
$(\lambda_F,\lambda_E)$ stage 1 & $(200,\,100)$ & $(1000,\,10)$ & $(200,\,100)$ \\
$(\lambda_F,\lambda_E)$ stage 2 & $(500,\,100)$ & $(1000,\,300)$ & $(500,\,100)$ \\
Batch size         & 64 & 32 & 64 \\
Learning rate      & $1.6\times10^{-2}$ & $10^{-3}$ & $1.6\times10^{-2}$ \\
Electrostatics     & n/a & disabled & disabled \\
\bottomrule
\end{tabular}
\end{table*}

The electrostatic term is disabled in every reported run, so the numbers
measure the local cluster expansion on electronic descriptors alone. The two
models are also matched in parameter count. Two confounds remain. MACE has no learning-rate warmup and holds its second-stage
learning rate at $8\times10^{-3}$, where \OrbA\ warms up over 15 epochs and
continues to decay on plateau. Every number is also a single run at a fixed
seed, so differences of a few percent are not resolved. The factors we draw
conclusions from, $3.0$ between \OrbA\ and the feature-free baseline, two
between a regular delta and the direct model, and three between \OrbA\ and
\OrbO, are large enough that seed variation is unlikely to account for them.

\subsection{Accuracy: direct and delta learning}
\label{sec:ani-accuracy}

Table~\ref{tab:accuracy} reports per-atom energy and per-component force mean
absolute errors on the test set, computed identically for all models on
absolute energies, with $E_0$ and any baseline added back to prediction and
reference alike.

\begin{table}[!htbp]
\centering
\caption{ANI-1x $\omega$B97X/6-31G$^*$ test accuracy, 442,000 frames (441,906
for xTB arms). Energy MAE in meV/atom, force MAE in meV/{\AA}; lower is
better. Best value in each column is in bold.}
\label{tab:accuracy}
\vspace{5pt}
\small
\begin{tabular}{llcc}
\toprule
Model & Learning scheme & $E$ MAE & $F$ MAE \\
\midrule
MACE  & direct                 &  9.61 & 146.3 \\
MACE  & regular (HF/STO-3G)    &  5.27 &  73.7 \\
MACE  & regular (GFN2-xTB)     &  5.00 &  68.8 \\
\midrule
\OrbO & direct                 &  9.80 & 124.5 \\
\OrbO & regular (HF/STO-3G)    &  5.13 &  59.5 \\
\OrbO & regular (GFN2-xTB)     &  4.99 &  59.2 \\
\OrbO & \texttt{mfdh} (HF)        & 10.63 & 134.7 \\
\OrbO & \texttt{mfdh} (xTB)       &  9.66 & 124.0 \\
\OrbO & \texttt{mfah} (HF)        & 10.99 & 148.8 \\
\OrbO & \texttt{mfah} (xTB)       &  9.31 & 120.2 \\
\midrule
\OrbA & direct                 &  3.16 &  76.1 \\
\OrbA & regular (HF/STO-3G)    &  \textbf{1.48} &  34.9 \\
\OrbA & regular (GFN2-xTB)     &  1.63 &  \textbf{31.7} \\
\OrbA & \texttt{mfdh} (HF)        &  2.99 &  73.3 \\
\OrbA & \texttt{mfdh} (xTB)       &  3.20 &  76.6 \\
\OrbA & \texttt{mfah} (HF)        &  3.37 &  81.3 \\
\OrbA & \texttt{mfah} (xTB)       &  3.54 &  85.3 \\
\bottomrule
\end{tabular}
\end{table}

\subsection{What the comparison shows}
\label{sec:ani-analysis}

First, \OrbA's electronic features help substantially, and in energy help more
than a delta baseline does. Trained directly, \OrbA\ reaches $3.16$\,meV/atom
and $76.1$\,meV/{\AA}, against $9.61$ and $146.3$ for the feature-free baseline
of the same architecture, a factor of $3.0$ in energy and $1.9$ in force.
MACE's most effective delta, a GFN2-xTB baseline, brings it to
$5.00$\,meV/atom, still $1.6$ times \OrbA's direct error, although in force both
delta-corrected MACE arms, at $73.7$ and $68.8$\,meV/{\AA}, sit marginally below
\OrbA's direct $76.1$. Featurizing the input with a cheap electronic-structure
calculation is therefore worth more than subtracting a comparably cheap one
from the output for energies, and about as effective for forces. On top of the
same baselines the features keep their advantage, \OrbA\ being $3.6$ and $3.1$
times more accurate in energy than MACE with an HF or xTB baseline and $2.1$ to
$2.2$ times in force. The orbital model does not share this advantage in
energy: \OrbO\ is level with MACE in every matched arm, at $9.80$ against
$9.61$\,meV/atom direct and $5.13$ and $4.99$ against $5.27$ and $5.00$ with a
regular baseline, and ahead only in force, at $124.5$ against
$146.3$\,meV/{\AA} direct.

Second, delta learning and electronic features compose: a regular baseline
improves \OrbA\ from $3.16$ to $1.48$\,meV/atom with HF/STO-3G and to $1.63$
with xTB, roughly halving the force error, so the two mechanisms address
different parts of the residual.

Third, and least expected, the atomic model beats the orbital model in every
matched arm, at $3.16$ against $9.80$\,meV/atom direct and $1.48$ against
$5.13$ with an HF baseline, despite \OrbO's strictly richer electronic
information. The configurations differ, \OrbO\ running 2 layers at
$r_{\max}=6$\,{\AA} and \OrbA\ 3 at $5$\,{\AA}, but at matched depth and width
on rMD17 (Sec.~\ref{sec:rmd17-orbital}) the gap does not close. The more
economical reading is that \OrbO's frozen-feature force path imposes an
accuracy ceiling, because the gradient it fits to reference forces cannot
contain the feature-response term of Eq.~\eqref{eq:orbforce-split}. The data
are equivocal on this: were the omitted term the whole story the gap should be
widest in force, yet on the direct arm \OrbO\ trails by a factor of $3.1$ in
energy and only $1.64$ in force, and the same ordering holds in every arm.

Fourth, \texttt{mfdh} does not reach the accuracy of a regular delta at this
data volume. Its \OrbA\ arms land at $2.99$ and $3.20$\,meV/atom, essentially
at the direct model's $3.16$ and about twice the regular arms; for \OrbO\ the
HF arm is slightly worse than direct and the xTB arm marginally better. The
multi-fidelity loss fits the absolute reference and only anchors a base head to
the cheap level, so it never gets the conditioning benefit of a smaller
residual target; what it buys is independence from the baseline at inference
(Sec.~\ref{sec:ani-baseline}). The parallel mode, designed to remove the
contention between \texttt{mfdh}'s two energy terms, does not recover the
deficit: for \OrbA\ it is worse than both the direct and the \texttt{mfdh} arm,
and for \OrbO\ only its xTB arm improves on direct. Since \texttt{mfah}'s
high-fidelity path is that of the direct model, this isolates the auxiliary
task itself: at $10{,}000$ frames, shaping the representation toward the cheap
level is not reliably helpful.

\subsection{Accuracy against baseline dependence at inference}
\label{sec:ani-baseline}

The arms are not interchangeable in deployment even when their errors are
similar. A \texttt{regular} arm must re-evaluate its baseline for every new
configuration (Eq.~\ref{eq:regular}), whereas the \texttt{direct},
\texttt{mfdh} and \texttt{mfah} arms evaluate nothing external.
Table~\ref{tab:baseline-dep} pairs each \OrbA\ arm's accuracy with that
requirement.

\begin{table*}[!tbp]
\centering
\caption{\OrbA\ accuracy against what each arm must evaluate for every new
configuration. The two regular arms buy their accuracy by reintroducing a
per-configuration calculation; the multi-fidelity arms need nothing external.}
\label{tab:baseline-dep}
\vspace{5pt}
\small
\begin{tabular}{lccl}
\toprule
Learning scheme & $E$ MAE [meV/atom] & $F$ MAE [meV/{\AA}] & Required per configuration \\
\midrule
regular (HF/STO-3G)  & 1.48 & 34.9 & a Hartree--Fock SCF \\
regular (GFN2-xTB)   & 1.63 & 31.7 & a GFN2-xTB evaluation \\
\texttt{mfdh} (HF)  & 2.99 & 73.3 & nothing (learned base head) \\
direct               & 3.16 & 76.1 & nothing \\
\texttt{mfdh} (xTB) & 3.20 & 76.6 & nothing (learned base head) \\
\texttt{mfah} (HF)  & 3.37 & 81.3 & nothing (cheap head discarded) \\
\texttt{mfah} (xTB) & 3.54 & 85.3 & nothing (cheap head discarded) \\
\bottomrule
\end{tabular}
\end{table*}

The most accurate arm, regular HF at $1.48$\,meV/atom, owes its factor of
$2.1$ over the direct model to a self-consistent field solution for every
configuration, exactly the calculation \OrbA's featurizer exists to avoid,
whereas the same baseline is nearly free for \OrbO, whose features already
require the SCF. The xTB arm, at $1.63$\,meV/atom for a cheap evaluation, is the
more attractive regular option but remains an external dependency.

The one combination that fails the precondition of Sec.~\ref{sec:mf} bears it
out. \OrbO's \texttt{mfdh} and \texttt{mfah} HF arms reach $10.63$ and
$10.99$\,meV/atom where its regular HF arm reaches $5.13$ at no additional
inference cost, so those rows are a configuration with no rationale rather than
evidence against the multi-fidelity schemes.\footnote{The argument concerns
the \emph{energy}. A Hartree--Fock baseline force is not free, since the
analytic gradient needs derivative integrals and must be taken before
localization (Remark~\ref{rem:hfgrad}), so the saving a learned head offers on
\OrbO\ is partial rather than nil, but far smaller than on \OrbA.} \OrbA's
\texttt{mfdh} HF arm is level with the direct model, at $2.99$ against
$3.16$\,meV/atom and $73.3$ against $76.1$\,meV/{\AA}, and adds an
interpretable decomposition into a learned cheap-level surrogate plus a
correction, while both \texttt{mfah} arms fall below the direct model at the
same requirement. At $10{,}000$ frames, then, the cheap level helps \OrbA\ when
it enters the energy and not when it only shapes the representation;
Sec.~\ref{sec:mf-scale} shows how this changes with more cheap data.

\subsection{Scaling the cheap level}
\label{sec:mf-scale}

Every multi-fidelity number so far was trained at label density $\rho = 1$, on
$10{,}000$ frames carrying both levels. That is not the regime the schemes were
designed for: the cheap level costs a small fraction of the reference, so the
natural dataset runs it on far more configurations, with the objective of
Eq.~\eqref{eq:mf-partial}. Table~\ref{tab:mf-scale} extends the cheap level to
$50{,}000$ frames with the high-fidelity subset held at $10{,}000$, so that four
frames in five carry a cheap label and no reference, and changes nothing else.

\begin{table*}[!tbp]
\centering
\caption{ANI-1x, \OrbA, effect of extending the cheap level from $10{,}000$
frames to $50{,}000$, of which $10{,}000$ carry a high-fidelity label
throughout. Energy MAE in meV/atom, force MAE in meV/{\AA}. The direct and
regular arms are repeated from Table~\ref{tab:accuracy} for reference and are
unchanged, since neither consumes the unlabelled frames.}
\label{tab:mf-scale}
\vspace{5pt}
\small
\begin{tabular}{lrrrr}
\toprule
 & \multicolumn{2}{c}{cheap level on 10\,k}
 & \multicolumn{2}{c}{cheap level on 50\,k} \\
\cmidrule(lr){2-3}\cmidrule(lr){4-5}
Learning scheme & $E$ & $F$ & $E$ & $F$ \\
\midrule
direct              & 3.16 & 76.1 & \multicolumn{2}{c}{---} \\
regular HF          & 1.48 & 34.9 & \multicolumn{2}{c}{---} \\
regular xTB         & 1.63 & 31.7 & \multicolumn{2}{c}{---} \\
\midrule
\texttt{mfdh} HF    & 2.99 & 73.3 & \textbf{1.59} & \textbf{37.6} \\
\texttt{mfdh} xTB   & 3.20 & 76.6 & 1.73 & 40.3 \\
\texttt{mfah} HF    & 3.37 & 81.3 & 3.10 & 71.4 \\
\texttt{mfah} xTB   & 3.54 & 85.3 & 3.14 & 80.3 \\
\bottomrule
\end{tabular}
\end{table*}

The additive mode is transformed. \texttt{mfdh} on a Hartree--Fock cheap level
goes from $2.99$ to $1.59$\,meV/atom and from $73.3$ to $37.6$\,meV/{\AA}, the
xTB arm behaving the same way, which puts it within eight percent of regular HF
in both channels while evaluating nothing outside the network. The parallel
mode improves by only six to twelve percent, to $3.10$ and $71.4$ on
Hartree--Fock and $3.14$ and $80.3$ on xTB, which leaves it beside the direct
model. The asymmetry follows from where the head sits: under \texttt{mfdh} a
cheap frame supervises an addend of the reported energy, while under
\texttt{mfah} it refines features the high-fidelity readout is free to ignore.
The rMD17 reversal, where \texttt{mfdh} was the worse mode at $950$ frames with
no extra cheap data, is consistent with this: \texttt{mfdh} suffers when the
cheap level is undersupplied and benefits when it is not.

\subsection{Single-molecule surfaces: revised MD17}
\label{sec:rmd17}

The revised MD17 set~\cite{rmd17} tests the opposite regime: one model per
molecule, trained on 950 configurations with 50 for validation and 1000 for
test, against PBE/def2-SVP labels recomputed on the MD17
trajectories~\cite{md17}. With this little data an electronic prior should
show if anywhere. Following this literature, energies are total rather than
per-atom MAE; Tables~\ref{tab:rmd17} and~\ref{tab:rmd17-forces} report all ten
molecules for all seven learning schemes, beside the published MACE results.

\begin{table*}[!tbp]
\centering
\caption{Revised MD17, one model per molecule trained on 950 configurations
and validated on 50. Total energy MAE in meV; lower is better. The MACE column
is the published result of Ref.~\cite{mace} (two layers, 256 channels,
$\ell_{\max}=3$, same 950/50 protocol); all other columns are \OrbA, evaluated
on 1000 configurations, with column heads abbreviating the learning schemes of
Table~\ref{tab:low-schemes}, \texttt{dh} for \texttt{mfdh} and \texttt{ah} for
\texttt{mfah}.}
\label{tab:rmd17}
\vspace{5pt}
\footnotesize
\setlength{\tabcolsep}{5pt}
\begin{tabular}{lrrrrrrrr}
\toprule
molecule & MACE & direct & \shortstack{reg\\HF} & \shortstack{reg\\xTB}
 & \shortstack{\texttt{dh}\\HF} & \shortstack{\texttt{dh}\\xTB}
 & \shortstack{\texttt{ah}\\HF} & \shortstack{\texttt{ah}\\xTB} \\
\midrule
aspirin       & 2.2 & 5.34 & 3.19 & 1.87 & 5.98 & 5.92 & 3.64 & 3.00 \\
azobenzene    & 1.2 & 3.37 & 1.93 & 0.87 & 3.71 & 3.68 & 2.23 & 2.15 \\
benzene       & 0.4 & 0.34 & 0.38 & 0.09 & 0.52 & 0.62 & 2.75 & 1.83 \\
ethanol       & 0.4 & 1.46 & 0.97 & 0.45 & 1.51 & 1.47 & 1.03 & 1.11 \\
malonaldehyde & 0.8 & 2.39 & 1.57 & 0.65 & 2.62 & 2.54 & 1.31 & 1.33 \\
naphthalene   & 0.5 & 1.81 & 0.91 & 0.34 & 2.04 & 1.97 & 5.69 & 3.29 \\
paracetamol   & 1.3 & 4.51 & 2.16 & 1.10 & 4.64 & 4.84 & 2.36 & 3.34 \\
salicylic     & 0.9 & 2.35 & 1.57 & 0.62 & 2.56 & 2.56 & 2.23 & 1.61 \\
toluene       & 0.5 & 1.65 & 0.68 & 0.35 & 1.81 & 1.85 & 1.30 & 1.68 \\
uracil        & 0.5 & 1.65 & 1.06 & 0.46 & 1.90 & 1.72 & 0.88 & 0.87 \\
\midrule
mean          & 0.87 & 2.49 & 1.44 & \textbf{0.68} & 2.73 & 2.72 & 2.34 & 2.02 \\
\bottomrule
\end{tabular}
\end{table*}

\begin{table*}[!tbp]
\centering
\caption{Revised MD17, force MAE in meV/{\AA} on the same models and splits
as Table~\ref{tab:rmd17}; lower is better. Columns as there.}
\label{tab:rmd17-forces}
\vspace{5pt}
\footnotesize
\setlength{\tabcolsep}{5pt}
\begin{tabular}{lrrrrrrrr}
\toprule
molecule & MACE & direct & \shortstack{reg\\HF} & \shortstack{reg\\xTB}
 & \shortstack{\texttt{dh}\\HF} & \shortstack{\texttt{dh}\\xTB}
 & \shortstack{\texttt{ah}\\HF} & \shortstack{\texttt{ah}\\xTB} \\
\midrule
aspirin       & 6.6 & 13.8 & 8.5 & 5.0 & 14.2 & 14.4 & 8.1 & 8.4 \\
azobenzene    & 3.0 &  9.2 & 5.6 & 2.9 & 10.4 & 10.7 & 6.7 & 4.2 \\
benzene       & 0.3 &  2.6 & 1.3 & 1.0 &  3.0 &  2.7 & 2.4 & 2.6 \\
ethanol       & 2.1 &  7.9 & 5.0 & 2.8 &  8.1 &  7.8 & 5.8 & 5.6 \\
malonaldehyde & 4.1 & 11.1 & 6.9 & 3.8 & 11.5 & 11.4 & 7.5 & 7.5 \\
naphthalene   & 1.6 &  6.7 & 3.9 & 1.6 &  7.5 &  7.1 & 3.8 & 2.4 \\
paracetamol   & 4.8 & 11.9 & 6.7 & 3.8 & 12.2 & 12.5 & 6.0 & 6.2 \\
salicylic     & 3.1 & 10.6 & 7.2 & 3.1 & 10.9 & 11.1 & 5.7 & 5.9 \\
toluene       & 1.5 &  6.8 & 3.3 & 1.7 &  7.0 &  7.5 & 2.3 & 2.6 \\
uracil        & 2.1 &  9.6 & 6.4 & 2.8 & 10.7 & 10.4 & 4.5 & 4.5 \\
\midrule
mean          & \textbf{2.9} &  9.0 & 5.5 & \textbf{2.9} &  9.6 &  9.6 & 5.3 & 5.0 \\
\bottomrule
\end{tabular}
\end{table*}

Both regular schemes again beat the direct model by a wide margin, but change
places: here xTB, not Hartree--Fock, is the stronger baseline in energy, by a
factor of two. Regular xTB is best at $0.68$\,meV and $2.9$\,meV/{\AA} over the
ten molecules, regular HF second at $1.44$ and $5.5$, and the direct and both
\texttt{mfdh} arms cluster at $2.5$ to $2.7$\,meV and $9.0$ to
$9.6$\,meV/{\AA}. The gain from xTB is uniform, the ratio of direct to
regular-xTB energy error running from $2.9$ on aspirin to $5.3$ on naphthalene;
the one exception to the benefit of a baseline is benzene, where HF is
marginally worse than none in energy, $0.38$ against $0.34$\,meV, though better
in force.

Against the published MACE results~\cite{mace}, from a larger two-layer model
with $\ell_{\max}=3$ and 256 channels trained on the same 950 configurations,
\OrbA's direct model is less accurate on every molecule except benzene in
energy, by a factor of $2.9$ in mean energy, $2.49$ against $0.87$\,meV, and
$3.1$ in mean force, $9.0$ against $2.9$\,meV/{\AA}. Only regular xTB reaches
that level: it is ahead in mean energy, at $0.68$\,meV, and lower on nine of the
ten molecules, and level in mean force at $2.9$\,meV/{\AA}, lower on four
molecules, equal on two and higher on four, at the cost of an xTB evaluation
per configuration that MACE does not need. On ANI-1x the recipe-matched MACE
was the less accurate model by a factor of $3.0$ in energy; on rMD17 the
comparison is between different configurations, and at 950 configurations of
one molecule the electronic features do not by themselves close the gap to the
stronger feature-free architecture.

The \texttt{mfah} arms edge past the direct model in mean energy, at $2.34$ and
$2.02$ against $2.49$\,meV, and improve its forces substantially, at $5.3$ and
$5.0$ against $9.0$\,meV/{\AA}, but less consistently per molecule: the HF arm
is an order of magnitude worse on benzene, $2.75$ against $0.34$\,meV, and three
times worse on naphthalene, $5.69$ against $1.81$, with forces close to the
direct model's, the signature of a per-molecule energy offset rather than a
distorted surface. Both \texttt{mfah} arms beat both \texttt{mfdh} arms in the
mean, at $2.34$ and $2.02$ against $2.73$ and $2.72$\,meV, \emph{reversing} the
ANI-1x ranking of the two modes. We attribute this to data volume:
\texttt{mfdh}'s base head is an extra function to fit from the same labels,
affordable with $10{,}000$ diverse configurations but capacity spent on an
unscored quantity with $950$ configurations of one molecule, whereas
\texttt{mfah} keeps its head out of the total and degrades gracefully toward
the direct model.

\OrbO\ was also trained on all ten molecules at the same depth and width as
\OrbA\ and on the same splits. It trails \OrbA\ in every learning scheme, by a
factor of $28$ in energy and $6$ in force for the direct model, while its own
schemes order as \OrbA's do (Sec.~\ref{sec:rmd17-orbital}).

\subsection{Thermal extrapolation: 3BPA}
\label{sec:3bpa}

The 3BPA set~\cite{threebpa} tests extrapolation: the flexible molecule
3-(benzyloxy)pyridin-2-amine, labelled at $\omega$B97X/6-31G(d), trained on 450
configurations sampled at 300\,K with 50 for validation, and evaluated at 300,
600, and 1200\,K on 1669, 2138, and 2139 frames.

\begin{table*}[!tbp]
\centering
\caption{3BPA, \OrbA, trained at 300\,K and evaluated at three temperatures.
Total energy MAE $E$ in meV and force MAE $F$ in meV/{\AA}. The final column is
the ratio of force MAE at 1200\,K to that at 300\,K.}
\label{tab:3bpa}
\vspace{5pt}
\small
\begin{tabular}{lrrrrrrr}
\toprule
 & \multicolumn{2}{c}{300\,K} & \multicolumn{2}{c}{600\,K}
 & \multicolumn{2}{c}{1200\,K} & \\
\cmidrule(lr){2-3}\cmidrule(lr){4-5}\cmidrule(lr){6-7}
Learning scheme & $E$ & $F$ & $E$ & $F$ & $E$ & $F$ & $F$ ratio \\
\midrule
direct       & 5.30 & 12.9 & 14.89 & 25.8 & 56.34 & 65.8 & 5.1 \\
regular HF   & 2.80 &  5.8 &  5.63 & 11.6 & 17.71 & 30.2 & 5.2 \\
regular xTB  & \textbf{1.73} &  \textbf{4.5} &  \textbf{4.41} &  \textbf{9.5} & \textbf{12.27} & \textbf{21.5} & 4.8 \\
\texttt{mfdh} HF   & 5.57 & 12.0 & 12.07 & 22.9 & 31.40 & 52.7 & 4.4 \\
\texttt{mfdh} xTB  & 5.83 & 12.4 & 12.91 & 23.9 & 33.41 & 55.7 & 4.5 \\
\texttt{mfah} HF   & 8.70 & 12.0 & 13.24 & 22.6 & 28.98 & 54.1 & 4.5 \\
\texttt{mfah} xTB  & 8.80 & 12.9 & 13.60 & 24.1 & 30.75 & 57.1 & 4.4 \\
\bottomrule
\end{tabular}
\end{table*}

In force the electronic prior buys accuracy, not robustness. Regular xTB is
best at every temperature, but the degradation from 300 to 1200\,K lies between
$4.4$ and $5.2$ for all seven arms, so a baseline that cuts the force error
nearly threefold at 300\,K cuts it by nearly the same factor at 1200\,K. In
energy the direct model degrades by a factor of $10.6$, whereas every arm
carrying a baseline or a cheap head degrades by $3.3$ to $7.1$; the two
\texttt{mfah} arms are flattest, at $3.3$ and $3.5$, largely because their
300\,K errors are the largest, $8.70$ and $8.80$ against the direct model's
$5.30$. The two regular arms thus improve the error at every temperature and
slow its growth, while the learned heads slow the growth from a 300\,K error at
or above the direct model's, consistent with a baseline supplying a physically
correct energy scale at geometries the network has never seen.

The published MACE results on this benchmark~\cite{mace} are root-mean-square
errors, so Table~\ref{tab:3bpa-rmse} compares them with the root-mean-square
errors of the models of Table~\ref{tab:3bpa}. The direct arm trails MACE by a
factor of $2.1$ to $2.6$ in energy and $1.9$ to $2.1$ in force. Regular xTB is
ahead of MACE at every temperature, its margin growing from $1.3$ at 300\,K to
$1.6$ in energy and $1.8$ in force at 1200\,K, and regular HF is ahead in force
at every temperature and in energy above 300\,K. The multi-fidelity arms all
trail MACE, by $1.3$ to $3.7$ in energy and $1.4$ to $2.1$ in force.

\begin{table*}[!tbp]
\centering
\caption{3BPA root-mean-square errors, total energy $E$ in meV and force $F$
in meV/{\AA}. The MACE row is the published $L=2$ result of Ref.~\cite{mace},
the mean of three runs; the other rows are the \OrbA\ models of
Table~\ref{tab:3bpa}.}
\label{tab:3bpa-rmse}
\vspace{5pt}
\small
\begin{tabular}{lrrrrrr}
\toprule
 & \multicolumn{2}{c}{300\,K} & \multicolumn{2}{c}{600\,K}
 & \multicolumn{2}{c}{1200\,K} \\
\cmidrule(lr){2-3}\cmidrule(lr){4-5}\cmidrule(lr){6-7}
Model & $E$ & $F$ & $E$ & $F$ & $E$ & $F$ \\
\midrule
MACE                       &  3.0 &  8.8 &  9.7 & 21.8 & 29.8 &  62.0 \\
\midrule
\OrbA\ direct              &  7.15 & 18.8 & 20.49 & 40.5 & 76.56 & 117.3 \\
\OrbA\ regular HF          &  3.67 &  8.6 &  7.73 & 18.3 & 25.87 &  61.2 \\
\OrbA\ regular xTB         & \textbf{2.34} & \textbf{6.8} & \textbf{6.56} & \textbf{14.9} & \textbf{18.12} & \textbf{35.3} \\
\OrbA\ \texttt{mfdh} HF    &  7.41 & 17.4 & 16.25 & 35.3 & 42.78 &  90.6 \\
\OrbA\ \texttt{mfdh} xTB   &  7.55 & 17.9 & 17.28 & 36.8 & 45.13 &  96.3 \\
\OrbA\ \texttt{mfah} HF    & 11.24 & 17.0 & 17.13 & 33.5 & 40.18 &  87.5 \\
\OrbA\ \texttt{mfah} xTB   & 11.08 & 18.2 & 17.85 & 35.7 & 42.11 &  91.8 \\
\bottomrule
\end{tabular}
\end{table*}


\section{Summary}
\label{sec:summary}

OrbACE builds equivariant cluster-expansion potentials on an explicit
electronic-structure calculation rather than on atomic numbers and coordinates
alone. \OrbO\ uses localized molecular orbitals as nodes, made differentiable
with respect to nuclei by the Mulliken-weighted, dominant-atom-anchored
centroids of Eqs.~\eqref{eq:mw-centroid}--\eqref{eq:anchor}; its features are
precomputed, so its force is the gradient at fixed features. \OrbA\ replaces
the self-consistent field with a one-shot extended-H\"uckel model that is a
closed-form function of the coordinates, so autograd returns an exact gradient
(Proposition~\ref{prop:conservative}). What makes this work is the
differentiation: every quantity in the chain is a spectral function of a
symmetric matrix, and only the Daleckii--Krein divided-difference form is
finite, basis-independent and exact at the degeneracies every symmetric
molecule supplies, whereas a broadened eigenvector derivative carries an error
that survives $\epsilon\to0$ and depends on the linear-algebra backend.

Three findings stand out. First, in energy, exposing electronic structure to
the network is worth more than subtracting comparable physics from its target.
\OrbA\ trained directly reaches $3.16$\,meV/atom on ANI-1x against $9.61$ for
an element-and-geometry model of the same architecture, and beats that model's
best delta-corrected arm, $5.00$\,meV/atom, in energy though not in force. The
two mechanisms compose, a regular baseline taking \OrbA\ to $1.48$\,meV/atom,
and the advantage of a regular baseline over the direct model replicates on ten
single-molecule surfaces and on extrapolation to four times the training
temperature, while the ordering of the two multi-fidelity modes and of the two
baselines does not replicate between ANI-1x and rMD17. The comparison with a feature-free model at matched architecture is
specific to ANI-1x: against the published, larger MACE configuration on rMD17
and 3BPA~\cite{mace}, \OrbA's direct model trails by a factor of two to three,
and only its regular xTB arm is level or ahead.

Second, the coarser representation wins. \OrbA\ is better than \OrbO\ in every
matched arm, by about a factor of three in energy on ANI-1x and by more on
rMD17 at matched depth and width, despite \OrbO's richer electronic
information. We offer as a hypothesis that \OrbO's frozen-feature force path
imposes an accuracy ceiling: the gap is narrower in force than in energy, which
an argument resting on the force path alone would not predict, and the matched
comparison equalizes depth and width but not parameter count or optimization
behaviour. It does rule out the simpler explanation that the two models were
merely run at different depth and cutoff.

Third, the value of a cheap level depends on where it enters. A regular
Hartree--Fock baseline is nearly free for \OrbO, which already runs an SCF, and
self-defeating for \OrbA, for which it reintroduces the calculation the
featurizer exists to avoid. A learned cheap head removes that dependence and
pays off when the cheap level is available on more configurations than the
reference: with a cheap level on $50{,}000$ frames against $10{,}000$ labelled
ones, \texttt{mfdh} reaches $1.59$\,meV/atom and $37.6$\,meV/{\AA}, within eight
percent of a regular Hartree--Fock delta while evaluating nothing external; at
equal data it is no better than the direct model.

\OrbO\ thus offers the richer representation at an SCF cost per evaluation,
with forces conservative only up to its frozen features; \OrbA\ offers a
coarser, atom-resolved representation for closed-shell main-group systems at
the cost of two small eigensolves, an $O(n^5)$ integral transform replaced by
dense $O(n^3)$ algebra, with forces that are exact gradients of its energy. On
the evidence here the second is the better instrument, and we suspect the
reason is the exactness of its derivative rather than the quality of its
descriptors, consistent with the broader finding that force models which are
not gradients of the reported energy are difficult to use reliably in
simulation~\cite{darkforces}.

\balance
These conclusions are drawn from neutral, closed-shell, non-periodic molecules
of H, C, N and O, the composition of all three benchmarks, which lies inside
the narrower featurizer domain of Sec.~\ref{sec:tb-limits}; the stress of
Eq.~\eqref{eq:stress} and the handling of periodic images are not exercised by
any reported result.

\vspace{1.5ex}
\noindent\begin{minipage}{\columnwidth}
\section*{Code availability}
The OrbACE source code is available at \url{https://github.com/PsiQ/orbace}
under the Academic Software License v1.0, which permits academic non-commercial
use. The repository includes the commands and settings for preparing the
ANI-1x, rMD17 and 3BPA data and for training and evaluating both models.
\end{minipage}

\vspace{1.5ex}
\noindent\begin{minipage}{\columnwidth}
\section*{Acknowledgments}
We thank Philipp T. Ernst and Sam Pallister for critically reviewing the manuscript.
We also thank the broader quantum chemistry team at PsiQuantum for helpful discussions and feedback.
\end{minipage}
\clearpage

\appendix
\nobalance
\setcounter{section}{0}
\renewcommand{\thesection}{S\arabic{section}}
\setcounter{table}{0}
\renewcommand{\thetable}{S\arabic{table}}
\setcounter{equation}{0}
\renewcommand{\theequation}{S\arabic{equation}}
\renewcommand{\theHsection}{S\arabic{section}}
\renewcommand{\theHtable}{S\arabic{table}}
\renewcommand{\theHequation}{S\arabic{equation}}

\begin{center}
{\Large\textbf{Supporting Information}}\\[0.4em]
\normalsize OrbACE: Orbital and Atomic Cluster Expansion
\end{center}

\noindent This appendix collects the implementation detail, the numerical
safeguards, the training and pipeline protocols, and the supplementary
tables that the main text refers to but does not need in order to be read.
Section, equation, and table numbers here are prefixed with S.

\section{Notation, units, and index conventions}
\label{si:notation}

\subsection{Units}

Internally, electronic quantities are kept in atomic units, with energies and
integrals in Hartree ($\Eh$), orbital moments and tight-binding coordinates in
Bohr ($a_0$). This is the scale on which they are conditioned and the
scale the underlying integral code produces. Geometry entering the graph is in
{\AA}ngstr\"om, and labels and predictions are in eV and eV/{\AA}. The
conversions used are $1\,a_0 = 0.529\,177\,210\,903$\,{\AA} and
$1\,\Eh = 27.211\,386\,245\,988$\,eV.

\subsection{Index conventions}

Uppercase Latin $A,B$ index atoms; lowercase $i,j$ index molecular orbitals or,
in Sec.~\ref{sec:atomic}, tight-binding eigenstates; Greek $\mu,\nu$ index
atomic orbitals. $Z_A$ is an atomic number and $\R_A$ a nuclear position;
$\rvec_i$ is an orbital centroid and $s_i$ an orbital species
(Eq.~\ref{eq:species}). A matrix written in the AO basis carries Greek indices
($\mathbf{F}_{\mu\nu}$) and its MO-basis transform carries Latin indices
($F_{ij}=[\mathbf{C}^\top\mathbf{F}\mathbf{C}]_{ij}$). For a matrix $\mathbf{M}$
in the AO basis, $\mathbf{M}_{AB}$ denotes the rectangular block of rows
belonging to atom $A$ and columns belonging to atom $B$, and $\mathbf{M}_{AA}$
the square on-site block. The Hadamard (elementwise) product is $\circ$, the
Frobenius norm $\|\cdot\|_F$, and $\mathcal{N}(i)$ the neighbor set of node $i$.

\section{Details of the shared equivariant backbone}
\label{si:backbone}

\subsection{Radial and angular edge embeddings}
\label{sec:radial}

Let $\mathbf{v}_{ij}$ be the displacement from node $i$ to node $j$ and
$r_{ij}=\|\mathbf{v}_{ij}\|$. The radial dependence is expanded in a Bessel
basis multiplied by a polynomial envelope,
\begin{equation}
  \begin{gathered}
  b_n(r) = \sqrt{\tfrac{2}{r_{\max}}}\;
           \frac{\sin\!\big(n\pi r/r_{\max}\big)}{r}\;\phi_{\mathrm{cut}}(r), \\
  n = 1,\dots,n_{\mathrm{bessel}},
  \end{gathered}
  \label{eq:bessel}
\end{equation}
with $n_{\mathrm{bessel}}=8$, and the envelope is the degree-$p$ polynomial of
Ref.~\cite{dimenet},
\begin{equation}
  \begin{aligned}
  \phi_{\mathrm{cut}}(r) &=
  \Big[1 - \tfrac12 (p{+}1)(p{+}2)\,x^{p}
        + p(p{+}2)\,x^{p+1} \\
  &\qquad - \tfrac12 p(p{+}1)\,x^{p+2}\Big]\,\mathbb{1}[x<1],
  \qquad x = \frac{r}{r_{\max}},
  \end{aligned}
  \label{eq:polycut}
\end{equation}
with $p=5$, which vanishes at $x=1$ with its first and second derivatives. The
angular dependence enters through real spherical harmonics
$Y_{\ell m}(\hat{\mathbf{v}}_{ij})$ to $\ell_{\max}=2$. Invariant chemical edge
features are projected into the radial space and added to
Eq.~\eqref{eq:bessel} after multiplication by the same envelope
(Secs.~\ref{sec:embed}, \ref{sec:atom-embed}).

\subsection{Interaction and product blocks}
\label{sec:msgpass}

Each of the $L$ layers applies an equivariant convolution followed by a
symmetric product. With $\mathbf{h}_i^{(t)}$ the node features at layer $t$,
$\mathbf{a}_i$ the one-hot species attribute, and $\mathbf{e}_{ij}$ the
invariant edge embedding,
\begin{align}
  \mathbf{m}_i^{(t)} &= \frac{1}{\bar{N}}\sum_{j\in\mathcal{N}(i)}
     \mathbf{R}^{(t)}\!\big(\mathbf{e}_{ij}\big)\,
     \Big(Y_{\ell}(\hat{\mathbf{v}}_{ij}) \otimes \mathbf{h}_j^{(t)}\Big),
     \label{eq:conv}\\[2pt]
  \mathbf{h}_i^{(t+1)} &=
     \mathrm{Product}^{(t)}\!\Big(\mathbf{m}_i^{(t)},\,
                                  \mathbf{h}_i^{(t)},\,\mathbf{a}_i\Big),
     \label{eq:product}
\end{align}
where $\otimes$ is the Clebsch--Gordan tensor product, $\mathbf{R}^{(t)}$ a
learnable perceptron that weights the tensor-product paths, and $\bar{N}$ a
fixed average neighbor count from the training set. The product block forms the
symmetrized tensor powers of the message up to correlation order $\nu$,
\begin{equation}
  \mathbf{m}_i,\quad
  \big(\mathbf{m}_i \otimes \mathbf{m}_i\big)_{\mathrm{sym}},\quad
  \dots,\quad
  \big(\underbrace{\mathbf{m}_i \otimes \cdots \otimes \mathbf{m}_i}_{\nu}\big)_{\mathrm{sym}},
\end{equation}
and contracts them into the target irreps. Correlation order $\nu$ resolves body
order $\nu+1$ within one layer, and stacking compounds this as
$B_t = \nu B_{t-1} + 1$ with $B_0 = 1$, giving
\begin{equation}
  B_L = \sum_{t=0}^{L} \nu^{t} = \frac{\nu^{L+1}-1}{\nu-1}
  \label{eq:bodyorder}
\end{equation}
in principle, a nominal $13$ for \OrbO's two layers and $40$ for \OrbA's three
at $\nu=3$, upper bounds since the product basis is truncated at $\ell_{\max}$.
The first layer is non-residual, later ones residual, and the last keeps only
scalar outputs. Table~\ref{tab:defaults} lists the defaults.

\begin{table*}[!tbp]
\centering
\caption{Default backbone hyperparameters.}
\label{tab:defaults}
\vspace{5pt}
\small
\renewcommand{\arraystretch}{1.1}
\begin{tabular}{lcc}
\toprule
 & \OrbO & \OrbA \\
\midrule
cutoff $r_{\max}$ [{\AA}]              & 6.0 & 5.0 \\
Bessel functions $n_{\mathrm{bessel}}$ & 8 & 8 \\
envelope degree $p$ (radial)           & 5 & 5 \\
maximum harmonic $\ell_{\max}$         & 2 & 2 \\
interaction layers $L$                 & 2 & 3 \\
hidden irreps                          & $64\times 0e + 64\times 1o + 64\times 2e$ & $64\times 0e + 64\times 1o + 64\times 2e$ \\
correlation order $\nu$                & 3 & 3 \\
radial MLP widths                      & $[64,64,64]$ & $[64,64,64]$ \\
ZBL envelope degree                    & 6 & 6 \\
\bottomrule
\end{tabular}
\end{table*}

\subsection{Readout and the gradient-detached scale/shift}
\label{sec:readout-shared}

Every layer contributes a per-node scalar through its own readout, linear for
the intermediate layers and a gated perceptron for the last, and the per-node
interaction energy is
\begin{equation}
  E^{\mathrm{int}}_i =
     \big(s_{\mathrm{scale}}\big)_{\mathrm{detach}}
     \sum_{t=1}^{L} \mathrm{Readout}^{(t)}\!\big(\mathbf{h}_i^{(t)}\big)
     \;+\; s_{\mathrm{shift}},
  \label{eq:readout}
\end{equation}
where $s_{\mathrm{scale}}$, the training-set root-mean-square force, and
$s_{\mathrm{shift}}$ are training-set statistics (Sec.~\ref{sec:stats}) held
constant in the backward pass.

\subsection{Short-range pair repulsion}
\label{sec:zbl}

Both models add a fixed screened-Coulomb repulsion of
Ziegler--Biersack--Littmark form~\cite{zbl} on \emph{atom} pairs, including in
\OrbO, whose message-passing graph is not atom-based:
\begin{equation}
  \begin{gathered}
  E_{\mathrm{ZBL}} = \sum_{A<B}
     \frac{k\,Z_A Z_B}{r_{AB}}\;
     \varphi\!\left(\frac{r_{AB}}{a_{AB}}\right)\,
     f_{\mathrm{env}}(r_{AB}), \\
  k = 14.3996\ \mathrm{eV\,\text{\AA}},
  \end{gathered}
  \label{eq:zbl}
\end{equation}
with the universal four-term screening function
\begin{equation}
  \begin{gathered}
  \varphi(x) = \sum_{k=1}^{4} c_k\,e^{-d_k x}, \\
  \begin{aligned}
    \mathbf{c} &= (0.1818,\; 0.5099,\; 0.2802,\; 0.028\,17),\\
    \mathbf{d} &= (3.2,\; 0.9423,\; 0.4028,\; 0.2016),
  \end{aligned}
  \end{gathered}
  \end{equation}
and the screening length
\begin{equation}
  a_{AB} = \frac{0.4543 \times 0.529\ \text{\AA}}
                {Z_A^{0.300} + Z_B^{0.300}} .
  \label{eq:zbl-a}
\end{equation}
The prefactor and exponent are those of the reference implementation, frozen by
default, not the published universal screening length ($0.8854\,a_0$, exponent
$0.23$): they give $a_{AB} = 0.070$\,{\AA} for a carbon pair against
$0.155$\,{\AA}, a harder wall. The $\mathbf{c}$ and $\mathbf{d}$ coefficients
are the published fit. The envelope $f_{\mathrm{env}}$ is
Eq.~\eqref{eq:polycut} with degree $6$ against a pair cutoff equal to the sum of
covalent radii, confining the term to core repulsion.


\section{Details of the orbital model}
\label{si:orbital}

\subsection{The self-consistent field}
\label{sec:scf}

The SCF uses PySCF~\cite{pyscf} at a chosen Gaussian basis (STO-3G by default),
restricted Hartree--Fock for closed shells and restricted open-shell
Hartree--Fock otherwise, converged to $10^{-10}\,\Eh$ in at most $200$
iterations with a second-order solver as fallback, on CPU or on GPU through
\texttt{gpu4pyscf}. The basis is a \emph{featurization} basis, not an accuracy
basis: STO-3G is poor for an energy, but which orbital sits where, how strongly
two orbitals couple and how polarized a bond is are qualitatively correct and
smooth in the geometry. What matters is that the same basis and localizer are
used for data preparation, training and inference.

\subsection{Orbital localization}
\label{sec:localization}

Canonical orbitals are delocalized, not transferable, and change character
discontinuously at eigenvalue crossings, so \OrbO\ localizes them into cores,
bonds and lone pairs, which recur across molecules. The doubly occupied
($n_i>1.5$), singly occupied and virtual ($n_i<0.5$) blocks are localized
independently, by Foster--Boys~\cite{boys}, matched to \OrbO's moment-based
features, or by Pipek--Mezey~\cite{pm}, which preserves $\sigma/\pi$
separation.

A localized set is not a Fock eigenbasis, so the orbital energy is redefined as
the diagonal of the Fock matrix rebuilt from the rotation-invariant $\mathbf{P}$,
\begin{equation}
  F_{ii} = \big[\mathbf{C}_{\mathrm{loc}}^\top\,\mathbf{F}\,
                \mathbf{C}_{\mathrm{loc}}\big]_{ii},
  \qquad
  \mathbf{F} = \mathbf{h} + \mathbf{v}_{\mathrm{HF}}[\mathbf{P}] ,
  \label{eq:fii}
\end{equation}
which becomes scalar node feature $0$; the off-diagonal $F_{ij}$, zero in the
canonical basis, are the couplings carried on the edges. Along a trajectory the
localizer can be warm-started from the previous coefficients projected into the
current occupied space and L\"owdin-orthonormalized,
\begin{equation}
  \begin{gathered}
  \mathbf{C}_0 = \mathbf{C}_{\mathrm{guess}}\,\mathbf{G}^{-1/2}, \\
  \mathbf{C}_{\mathrm{guess}} = \mathbf{C}\,\mathbf{C}^\top \mathbf{S}\,\mathbf{C}_{\mathrm{prev}}, \\
  \mathbf{G} = \mathbf{C}_{\mathrm{guess}}^\top \mathbf{S}\,\mathbf{C}_{\mathrm{guess}},
  \end{gathered}
  \label{eq:warmloc}
\end{equation}

\subsection{Orbital species}
\label{sec:species}

In place of the atomic number, \OrbO\ labels each orbital by a collision-free
encoding of (dominant element, dominant angular momentum, occupancy class),
\begin{equation}
  s_i \;=\; Z^{\mathrm{dom}}_i \cdot 15
        \;+\; \ell^{\mathrm{dom}}_i \cdot 3
        \;+\; c_i ,
  \label{eq:species}
\end{equation}
with $A^{\mathrm{dom}}_i = \arg\max_A |w_{iA}|$ from the weights of
Eq.~\eqref{eq:mulliken}, $\ell^{\mathrm{dom}}_i$ the angular channel carrying
the largest squared-coefficient mass (channels $\ell\ge 4$ folded into $g$), and
$c_i\in\{0,1,2\}$ for virtual, singly and doubly occupied orbitals. The stride
$15$ is the product of the two inner cardinalities, so triples never collide;
the table is sized to $\max_i s_i + 1$ for the dataset, since clamped indices
would silently merge orbital types.

\subsection{Scalar node features}
\label{sec:scalarfeat}

Each orbital carries the ten scalars of Table~\ref{tab:orbscalars}, each a
diagonal expectation value in the localized MO basis against the appropriate AO
integral matrix, with $\langle r^2\rangle_i$ referred to the molecular centre of
geometry by Eq.~\eqref{eq:r2-centered}.

\begin{table*}[!tbp]
\centering
\caption{The ten scalar node features of \OrbO.}
\label{tab:orbscalars}
\vspace{5pt}
\small
\renewcommand{\arraystretch}{1.1}
\begin{tabular}{clll}
\toprule
index & symbol & description & units \\
\midrule
0    & $F_{ii}$                          & Fock diagonal (orbital energy) in the localized basis & $\Eh$ \\
1    & $n_i$                             & occupancy, $\in\{0,1,2\}$ & dimensionless \\
2    & $h_{ii}$                          & core-Hamiltonian diagonal & $\Eh$ \\
3    & $g_{iiii}$                        & two-electron self-repulsion $(ii|ii)$ & $\Eh$ \\
4    & $\langle r^2\rangle_i$            & spatial second moment about $\bar{\R}$, Eq.~\eqref{eq:r2-centered} & $a_0^2$ \\
5    & $\tfrac12 n_i (h_{ii}+F_{ii})$    & orbital partition energy & $\Eh$ \\
6    & $\sigma_i$                        & orbital spread & $a_0$ \\
7--9 & $\lambda_1\!\ge\!\lambda_2\!\ge\!\lambda_3$ & quadrupole eigenvalues & $a_0^2$ \\
\bottomrule
\end{tabular}
\end{table*}

Feature 5 satisfies
\begin{equation}
  \sum_i \tfrac12 n_i\,(h_{ii}+F_{ii}) = E_{\mathrm{elec}},
  \label{eq:mullikenpartition}
\end{equation}
the exact Hartree--Fock electronic energy. The network does not receive the raw
value, which would let a readout reproduce the electronic energy by summation:
feature 5 belongs to group A of Sec.~\ref{sec:embed}, normalized per orbital in
a way invariant to rescaling and shifting. Features 6--9 derive from the
orbital second-moment tensor
\begin{equation}
  Q^{ab}_i = \mel{\psi_i}{r_a r_b}{\psi_i}
             - \rvec_{i,a}\,\rvec_{i,b},
  \label{eq:quad}
\end{equation}
the covariance of the electron position within orbital $i$, whose trace gives
the spread
\begin{equation}
  \sigma_i = \sqrt{\Tr \mathbf{Q}_i}
           = \sqrt{\langle r^2\rangle_i - |\rvec_i|^2}
  \label{eq:spread}
\end{equation}
and whose eigenvalues describe shape. Two auxiliary groups complete the
$30$-dimensional input: a binary 15-dimensional encoding of which shells
$1s,\dots,5g$ contribute, and the squared coefficient mass in each of the 5
angular channels.

\subsection{Edge features}
\label{sec:edge-feats}

Each edge carries the six MO-basis matrix elements of Table~\ref{tab:orbedges}.

\begin{table*}[!tbp]
\centering
\caption{The six chemical edge features of \OrbO.}
\label{tab:orbedges}
\vspace{5pt}
\small
\renewcommand{\arraystretch}{1.1}
\begin{tabular}{lll}
\toprule
quantity & definition & interpretation \\
\midrule
$F_{ij}$      & $[\mathbf{C}^\top\mathbf{F}\mathbf{C}]_{ij}$ & Fock coupling: direct orbital interaction strength \\
$J_{ij}$      & $(ii|jj)$ & Coulomb: classical density--density repulsion \\
$K_{ij}$      & $(ij|ji)$ & exchange: same-spin quantum interaction \\
$h_{ij}$      & $[\mathbf{C}^\top\mathbf{h}\mathbf{C}]_{ij}$ & core-Hamiltonian coupling, isolating one-electron effects \\
$|\Delta\varepsilon_{ij}|$ & $|F_{ii}-F_{jj}|$ & energy gap: the perturbation-theory denominator \\
$S_{ij}$      & $[\mathbf{C}^\top\mathbf{S}\mathbf{C}]_{ij}$ & MO overlap: $\delta_{ij}$ for the orthonormal localized set \\
\bottomrule
\end{tabular}
\end{table*}

Pipek--Mezey and Foster--Boys localization rotate the occupied and virtual
blocks unitarily, so the localized orbitals remain orthonormal and $S_{ij}$
vanishes on every edge (below $10^{-13}$ in practice): the channel carries no
information, and the independent content is five couplings per edge. $J$, $K$
and $g_{iiii}$ come from the single $O(n^5)$ two-electron transformation,
materialized only on graph edges. Each localized orbital is defined only up to
sign, and $F_{ij}$ and $h_{ij}$ change sign with either orbital, so the network
receives their magnitudes; $J_{ij}$, $K_{ij}$ and $|\Delta\varepsilon_{ij}|$ are
unaffected by the signs.

\subsection{Neighbor list, degenerate edges, and window pruning}
\label{sec:orbgraph}

Edges connect orbitals whose stored centroids lie within $r_{\max}$, while edge
vectors are rebuilt from the anchored positions in every forward pass. Orbitals
sharing a dominant atom can coincide after anchoring, so edges shorter than
$10^{-4}$\,{\AA} are masked \emph{before} any length or harmonic is evaluated;
masking afterwards gives $0\cdot\mathrm{NaN}=\mathrm{NaN}$ in the backward pass,
which the global $s_{\mathrm{spread}}$ would spread across the batch.
Optionally, only the $w$ highest occupied and $w$ lowest virtual orbitals are
retained; deep cores are absorbed by the per-species baseline, and pruning them
shrinks organic graphs three- to fivefold.

The dominant atom is a discrete label of a continuous quantity. Writing
Eq.~\eqref{eq:anchor} as
$s_{\mathrm{spread}}\tilde{\rvec}_i+(1-s_{\mathrm{spread}})\R_{A^{\mathrm{dom}}_i}$
shows that a switch of dominant atom moves the node by $(1-s_{\mathrm{spread}})$
times the bond vector; the polarization, referred to the dominant nucleus,
jumps by the whole bond vector, and if the two atoms differ in element the
species changes too. The jump in position vanishes only at
$s_{\mathrm{spread}}=1$, whereas the trained models have $s_{\mathrm{spread}}$
between $0.89$ and $1.04$. In the rMD17 training frames the two largest weights
lie within $0.02$ of each other for 38\% of benzene's orbitals and for 9\% and
11\% of those of ethanol and aspirin, almost all in bonds between like atoms.

\subsection{Embeddings}
\label{sec:embed}

The orbital species is one-hot encoded as the node attribute $\mathbf{a}_i$ and
passed through an equivariant linear layer to seed the scalar node features.
The $30$-dimensional scalar input is split into four groups, each normalized
separately by a layer normalization before fusion (Table~\ref{tab:orbgroups}).
\begin{table*}[!tbp]
\centering
\caption{The four normalization groups of \OrbO's scalar node input.}
\label{tab:orbgroups}
\vspace{5pt}
\small
\renewcommand{\arraystretch}{1.1}
\begin{tabular}{clll}
\toprule
group & content & dimension & scale \\
\midrule
A & orbital energies, occupancy, integrals, $\langle r^2\rangle_i$ (features 0--5) & 6  & mixed: $\Eh$, $a_0^2$, dimensionless \\
B & AO shell encoding                              & 15 & $\{0,1\}$ \\
C & per-$\ell$ coefficient norms                   & 5  & $[0,1]$, $\sum\approx 1$ \\
D & shape invariants (features 6--9)               & 4  & $a_0$, $a_0^2$, $O(1$--$10^2)$ \\
\bottomrule
\end{tabular}
\end{table*}
A single normalization would let group D set the scale for all thirty channels.
The groups pass through a SiLU perceptron of widths $(2m, m, m)$ whose output is
added to the species embedding, and a bounded, identity-centered occupation gate
then modulates the node features,
\begin{equation}
  \begin{gathered}
  g_i = 1 + m\,\tanh\!\big(\mathrm{MLP}(F_{ii}, n_i)\big), \\
  m = m_{\max}\,\sigma(\theta),
  \quad m_{\max} = 0.5,\ \ \theta = -2.2,
  \end{gathered}
  \label{eq:gate}
\end{equation}
with $\mathbf{h}_i \leftarrow g_i\,\mathbf{h}_i$ and a zero-initialized output
layer, so $g_i \equiv 1$ at initialization. Its layer-normalized input pair maps
to $\pm(1,-1)$, so the gate separates occupied from virtual orbitals but does
not see the orbital energy. The six edge features are fused with the radial
embedding through a base projection plus a bounded, zero-initialized residual,
\begin{equation}
  \begin{gathered}
  \mathbf{e}_{ij} = \mathbf{b}(r_{ij})
    + \Big[\mathrm{base}(\mathbf{c}_{ij})
           + \beta\,\mathrm{res}(\mathbf{c}_{ij})\Big]\,
      \phi_{\mathrm{cut}}(r_{ij}), \\
  \beta = \beta_{\max}\,\sigma(\theta),\ \ \beta_{\max}=0.5,
  \end{gathered}
  \label{eq:edgeembed}
\end{equation}
with $\mathbf{c}_{ij}$ the layer-normalized couplings and $\mathbf{b}$ the
Bessel embedding.

\subsection{Readout and pooling onto atoms}
\label{sec:orb-readout}

The per-orbital interaction energy follows Eq.~\eqref{eq:readout} with $i$
running over orbitals. In parallel, an atom energy head pools the scalar
channels of every interaction layer onto atoms with the Mulliken weights,
\begin{equation}
  \mathbf{x}_A = \sum_i w_{iA}\,
    \Big[\mathbf{h}_i^{(1)} \,\|\, \cdots \,\|\,
         \mathbf{h}_i^{(L)}\Big]_{\text{scalars}},
  \label{eq:atompool}
\end{equation}
and maps them through a shared SiLU perceptron to a correction
$E^{\mathrm{corr}}_A$ summed into $E_{\mathrm{corr}}$, a channel for naturally
atomic contributions such as those removed by window pruning.

\subsection{The frozen baseline}
\label{sec:e0-orb}

For \OrbO\ the species baseline of Eq.~\eqref{eq:energy-shared} is
\begin{equation}
  E_0 = \sum_i E^{0}_{s_i} \;+\; \sum_A E^{0,\mathrm{atom}}_{Z_A},
  \label{eq:e0-orb}
\end{equation}
with both terms frozen; the per-element term is fitted on element counts
(Sec.~\ref{sec:stats}) and the per-orbital-species table is zero by default. No
slope $b_s\,F_{ii}$ is included: $\sum_i F_{ii}$ double-counts
electron--electron repulsion, and the resulting geometry-dependent error would
sit in a baseline excluded from the forces.

\subsection{Long-range electrostatics}
\label{sec:longrange}

Both models offer an optional screened electrostatic term between graph sites,
disabled in every reported run. Each site carries a learned charge
$q_i = \mathrm{MLP}(\mathbf{h}_i)$ and, in \OrbO's multipole variant, a dipole
$\muvec_i = s_i\,\pvec_i$ along the frozen polarization of Eq.~\eqref{eq:pol}.
The energy uses Ewald-damped interaction tensors~\cite{ewald} with
$\alpha=0.2$; \OrbA\ uses nuclei as sites, monopole-only, and in both models
$E_{\mathrm{LR}}$ enters the differentiated energy but not the multi-fidelity
base head. The charge readout is initialized with weight standard deviation
$10^{-3}$: standard initialization sends the Coulomb energy to $10^7$\,eV,
while exact zero is a fixed point with zero gradient.


\section{Details of the atomic model}
\label{si:atomic}

\subsection{The minimal valence basis and the overlap matrix}
\label{sec:tb-overlap}

Each atom carries its valence shell only, so
$n_{\mathrm{AO}} = n_{\mathrm{H}} + 4\,n_{\mathrm{heavy}}$, and each atomic
orbital is a three-term contracted Cartesian Gaussian from the STO-3G valence
shells~\cite{hehre},
\begin{equation}
  \chi_\mu(\rvec) = \sum_{k=1}^{3} d_{\mu k}\,
     (x-A_x)^{l_x}(y-A_y)^{l_y}(z-A_z)^{l_z}\,
     e^{-\alpha_{\mu k}\|\rvec-\mathbf{A}\|^2},
  \label{eq:cgto}
\end{equation}
with $|\mathbf{l}| \in\{0,1\}$ and the contraction coefficients premultiplied by
the primitive normalization
$(2\alpha/\pi)^{3/4}$, times $(4\alpha)^{1/2}$ for a $p$ function. Period-2 and
period-3 elements each share one set of contraction coefficients, with
element-specific exponents shared between the $s$ and $p$ shells. With
$p = \alpha+\beta$, $\mu = \alpha\beta/p$, product center
$\mathbf{P} = (\alpha\mathbf{A}+\beta\mathbf{B})/p$,
$\mathbf{PA} = \mathbf{P}-\mathbf{A}$ and $\mathbf{PB} = \mathbf{P}-\mathbf{B}$,
the primitive overlap is
\begin{equation}
  S^{\mathrm{prim}} =
  \left(\frac{\pi}{p}\right)^{3/2}
  e^{-\mu\|\mathbf{A}-\mathbf{B}\|^2}
  \prod_{d\in\{x,y,z\}} S^{1\mathrm{d}}\big(l_d, m_d\big),
  \label{eq:primovlp}
\end{equation}
and since only $l_d,m_d\in\{0,1\}$ occur, the one-dimensional factors are
exhausted by four cases~\cite{mcmurchie},
\begin{equation}
  \begin{gathered}
  S^{1\mathrm{d}}(0,0) = 1, \\
  S^{1\mathrm{d}}(1,0) = PA_d, \\
  S^{1\mathrm{d}}(0,1) = PB_d, \\
  S^{1\mathrm{d}}(1,1) = PA_d\,PB_d + \frac{1}{2p},
  \end{gathered}
  \label{eq:s1d}
\end{equation}
selected arithmetically,
\begin{equation}
  \begin{aligned}
  S^{1\mathrm{d}} ={}&
    (1-l_d)(1-m_d)
    + l_d(1-m_d)\,PA_d\\
    &+ (1-l_d)m_d\,PB_d
    + l_d m_d\Big(PA_d\,PB_d + \tfrac{1}{2p}\Big),
  \end{aligned}
\end{equation}
so the whole overlap is a fixed sequence of elementwise tensor operations with
no recursion or branching. Contracting and renormalizing to unit self-overlap
gives
\begin{equation}
  \bar S_{\mu\nu} = \sum_{k,l} d_{\mu k} d_{\nu l}\,
                    S^{\mathrm{prim}}_{\mu k,\nu l},
  \qquad
  S_{\mu\nu} = \frac{\bar S_{\mu\nu}}
                    {\sqrt{\bar S_{\mu\mu}}\,\sqrt{\bar S_{\nu\nu}}},
  \label{eq:contracted-ovlp}
\end{equation}
so that $S_{\mu\mu}=1$ exactly, as the Wolfsberg--Helmholz form assumes.

\subsection{Redundancy in the feature layout}
\label{sec:tb-redundancy}

The twelve node and six edge channels are not independent: because
$\mathbf{F}\equiv\mathbf{H}$ (Eq.~\ref{eq:f-equals-h}), node features 7 and 8
coincide, as do edge features 2 and 4; node feature 11 equals feature 1 in a
valence-only basis; features 4--6 are identically zero; and feature 0 is affine
in feature 1 given the element, while features 2 and 3 sum to feature 1. The
independent content is five scalars per atom and five per edge; the zero
channels are harmless since the standardizer floors the standard deviation at
$10^{-6}$.

\subsection{Batched evaluation by block-diagonal padding}
\label{sec:tb-batching}

Solving one molecule at a time serializes many small double-precision
eigendecompositions, a stage bound by kernel-launch latency. \OrbA\ instead pads
molecules to a common size so that both matrices are exactly block diagonal,
with the padded spectrum held clear of the real one:
\begin{equation}
  \begin{gathered}
  \mathbf{S} =
  \begin{pmatrix}
    \mathbf{S}^{\mathrm{real}} & \mathbf{0}\\
    \mathbf{0} & s_{\mathrm{pad}}\mathbf{I}
  \end{pmatrix}, \\
  \mathbf{H} =
  \begin{pmatrix}
    \mathbf{H}^{\mathrm{real}} & \mathbf{0}\\
    \mathbf{0} & \varepsilon_{\mathrm{pad}}\mathbf{I}
  \end{pmatrix}, \\
  s_{\mathrm{pad}} = 10,\quad
  \varepsilon_{\mathrm{pad}} = 10^{3}\ \Eh .
  \end{gathered}
  \label{eq:padding}
\end{equation}
The real eigenpairs, features and gradients then agree with the per-molecule
path to floating-point precision. Padded slots have zero contraction
coefficients and unit exponents, so they contribute exactly zero while keeping
Eq.~\eqref{eq:primovlp} finite, and their levels sit at $100\,\Eh$, far above
every real H\"uckel level; identity padding would instead create
near-degeneracies with the renormalized overlap, whose eigenvalues cluster
around one. Bucketing batches by AO count reduces the padding overhead from
$3.38\times$ to $1.47\times$.

\subsection{Embeddings and dataset-level standardization}
\label{sec:atom-embed}

The atomic number is one-hot encoded and embedded as for \OrbO; the twelve
standardized node scalars pass through a SiLU perceptron of widths $(2m, m, m)$
added to the species embedding, and the six standardized edge features are
added to the Bessel embedding under the cutoff envelope,
\begin{equation}
  \mathbf{e}_{AB} = \mathbf{b}(r_{AB})
    + \mathrm{proj}\big(\hat{\mathbf{c}}_{AB}\big)\,
      \phi_{\mathrm{cut}}(r_{AB}),
  \label{eq:atom-edge-embed}
\end{equation}
with a zero-initialized output layer. The standardization is
\begin{equation}
  \hat{x}_k = \frac{x_k - \mu_k}{\sigma_k},
  \label{eq:standardize}
\end{equation}
with $\mu_k$ and $\sigma_k$ fixed buffers estimated once on up to 512 training
frames, not a per-sample normalization: a layer normalization would rescale each
atom's features by its own statistics, destroying the absolute magnitudes and
inter-atomic ratios of features 7--9 of Table~\ref{tab:tbnode} that carry the
physical signal. \OrbA\ has no occupation gate, since it has no orbital nodes to
weight.


\section{Equivariance of the orbital descriptors}
\label{si:equiv}

\subsection{The orbital model's descriptors}
\label{sec:equiv-orbital}

Since the molecular-orbital coefficients transform as
$\mathbf{C}\mapsto\mathbf{D}\mathbf{C}$, any molecular-orbital matrix element
inherits two cancelling factors,
\begin{equation}
  \mathbf{C}^\top\mathbf{M}\mathbf{C}
  \;\longmapsto\;
  (\mathbf{D}\mathbf{C})^\top
  \big(\mathbf{D}\mathbf{M}\mathbf{D}^\top\big)
  (\mathbf{D}\mathbf{C})
  = \mathbf{C}^\top\mathbf{M}\mathbf{C},
  \label{eq:mo-invariant}
\end{equation}
which covers $F_{ii}$, $h_{ii}$, and all six edge descriptors; the two-electron
quantities $J_{ij}$, $K_{ij}$ and $g_{iiii}$ carry four such factors. The
off-diagonal $F_{ij}$ and $h_{ij}$ are in addition odd under the sign
of either orbital, which the localizer does not fix, so they enter as magnitudes. The second
moment $\langle r^2\rangle_i$ needs care, because its operator carries an
origin. Referred to the origin of the input frame it is not translation
invariant, acquiring $2\mathbf{t}\cdot\rvec_i + |\mathbf{t}|^2$ under
$\R \mapsto \R + \mathbf{t}$. We therefore refer it to the molecular centre of
geometry $\bar{\R} = N_{\mathrm{at}}^{-1}\sum_A \R_A$,
\begin{equation}
  \langle r^2\rangle_i
    = \mel{\psi_i}{\,|\rvec - \bar{\R}|^2\,}{\psi_i} ,
  \label{eq:r2-centered}
\end{equation}
which translates with the molecule and restores both invariances. Centring on
the orbital's own centroid would collapse the channel onto $\sigma_i^2$, whereas
Eq.~\eqref{eq:r2-centered} equals $\sigma_i^2 + |\rvec_i - \bar{\R}|^2$ and
still separates a diffuse orbital from a compact one far from the centre. The
spread and the quadrupole eigenvalues are unaffected by the choice. Occupancies
are invariant by definition and the partition energy is a combination of
invariants; the per-$\ell$ coefficient norms are invariant because $\mathbf{D}$
is orthogonal within each shell, and the shell encoding records only which shell
types carry weight. The Mulliken weights of Eq.~\eqref{eq:mulliken} are
invariant by the block argument of Sec.~\ref{sec:ao-transform}, which matters
because they define the node positions.

The geometric quantities are carried in the channel their degree requires. The
centroid is the expectation of the position operator, so the anchored position
of Eqs.~\eqref{eq:mw-centroid}--\eqref{eq:anchor}, an affine combination of
nuclear positions with invariant coefficients summing to one, is a proper $1o$
object, which makes the orbital graph itself equivariant. The polarization is a
difference of two vectors and likewise $1o$, and the traceless quadrupole
transforms by congruence and is $2e$.

Two qualifications apply to \OrbO\ alone. The argument assumes the localizer
returns the image of the original localized set. 
OrbACE-O is not guaranteed to be invariant to distinct orbital-localization solutions, so switching between such solutions along a geometry trajectory could introduce artificial discontinuities in the predicted potential energy surface.
And because \OrbO's descriptors are
cached, the model is equivariant in geometry alone only because the featurizer
is itself equivariant and the inference path recomputes it; a rotated geometry
presented with cached, unrotated descriptors would give a wrong answer that
nothing in the model would detect.

Several natural constructions are avoided deliberately: raw coefficients
$C_{\mu i}$, which rotate with the AO index; per-orbital Mulliken populations
$\diag(\mathbf{P}\mathbf{S})$, which rotation redistributes within a shell,
hence the shell sums of Table~\ref{tab:tbnode}; Cartesian components of the
polarization or quadrupole placed in scalar channels; and the eigenvectors of
the quadrupole, which are frame-dependent and ill-defined within a degenerate
eigenspace, and whose orientational content the traceless quadrupole supplies
instead.


\section{Training protocol}
\label{si:training}

\subsection{Training-set statistics and the \texorpdfstring{$E_0$}{E0} regression}
\label{sec:stats}

Three statistics are computed on the training split alone and stored on the
checkpoint: the average neighbor count $\bar N$ that normalizes the message sum
of Eq.~\eqref{eq:conv}, the shift $s_{\mathrm{shift}}$, the mean per-node
interaction energy after $E_0$ subtraction, and the scale
$s_{\mathrm{scale}} = \varsigma$, the root-mean-square force component. The
per-element baseline is a least-squares regression of the total energies on
element counts, stored as a difference from the reference isolated-atom values,
\begin{equation}
  E^{0,\mathrm{atom}}_Z = E^{0,\mathrm{fit}}_Z - E^{0,\mathrm{ref}}_Z ,
  \label{eq:e0fit}
\end{equation}
with the references taken from the dataset's \texttt{IsolatedAtom} frames when
present and otherwise from single-atom Hartree--Fock at the feature basis; all
splits must share them, or predictions and references differ by a constant per
element.

\subsection{Optimization protocol}
\label{sec:optim}

Training uses Adam or AdamW with AMSGrad, a linear learning-rate warmup, and a
reduce-on-plateau schedule with a floor of $10^{-6}$. An exponential moving
average of the weights with decay $0.999$~\cite{ema}, activated after warmup
from the current weights, is used for validation, early stopping and
checkpoints, and stochastic weight averaging~\cite{swa} is available as a final
phase. The loss weights follow a two-stage schedule that raises the energy
weight once the forces are fitted, resetting the best-validation record and
early-stopping patience at the transition (weights in
Table~\ref{tab:runconfig}). Gradients are clipped at norm $10$, batches with
non-finite losses are skipped, and under bfloat16 autocast nuclear positions
stay in single precision. The trainer keeps the best-validation, best-force,
final and averaged models separately, since the best-force checkpoint rarely
coincides with the best-validation one. Because \OrbA's featurizer is
launch-bound (Sec.~\ref{sec:tb-batching}), its default batch size is $64$ with
the learning rate scaled by $\sqrt{64/16}$, unless a batch size is supplied. The
trade is small: at 1000 epochs \OrbA's direct arm reaches $3.162$\,meV/atom at
batch size 64 and $3.045$ at 16, a $3.8\%$ difference that a single seed does
not resolve.

\begin{table*}[!tbp]
\centering
\caption{Cost structure of the two models. $n_{\mathrm{MO}}$ and
$n_{\mathrm{AO}}$ are the retained orbital and valence AO counts.}
\label{tab:cost}
\vspace{5pt}
\small
\renewcommand{\arraystretch}{1.2}
\begin{tabular}{lll}
\toprule
 & \OrbO & \OrbA \\
\midrule
Featurizer                & self-consistent, iterative & closed form, one shot \\
Dominant featurizer term  & $O(n_{\mathrm{MO}}^5)$ integral transform & $O(n_{\mathrm{AO}}^3)$ eigensolve \\
Featurizer precision      & double (library) & double (required, see Sec.~\ref{sec:tb-solve}) \\
When paid                 & once per frame, offline & every forward pass \\
Graph nodes per atom      & several (fewer after pruning) & exactly one \\
Preparation cost          & dominant one-time cost & negligible \\
Per-step inference cost   & SCF $+$ localization, localization dominant & two small eigensolves \\
Bottleneck character      & compute-bound & launch-bound, hence batched \\
\bottomrule
\end{tabular}
\end{table*}

\section{Data and inference pipelines}
\label{sec:pipeline}

\subsection{Dataset curation by farthest-point sampling}

Training and validation splits are selected by farthest-point sampling in a
descriptor space, either a geometry mode (normalized composition plus sorted
Coulomb-matrix eigenvalues) or an orbital mode (per-frame mean and standard
deviation of the ten orbital scalars plus composition). Descriptors are standardized per dimension and
greedy selection repeatedly adds
\begin{equation}
  x_{k+1} = \arg\max_{x\notin\mathcal{S}}\ \min_{y\in\mathcal{S}}\ \|x-y\|^2 ,
  \label{eq:fps}
\end{equation}
from a random seed, at cost $O(N_{\mathrm{sel}} N_{\mathrm{tot}})$ with a
running minimum-distance array. The training and validation subsets are
carved from the selection, the remainder becomes the test set, and
isolated-atom frames are propagated to every subset.

\subsection{Feature precomputation, and its absence}

For \OrbO, a parallel preparation step runs the SCF, localization,
featurization, an optional baseline and the $E_0$ subtraction per frame, and
writes per-frame archives whose metadata record the basis, localizer and
baseline. The features are baseline-independent
(Eq.~\ref{eq:delta-targets}), so baselines are attached by cheap
re-baselining passes, and a Hartree--Fock baseline is free because the
preparation pass already stored its energy and forces. For \OrbA\ there is
nothing electronic to precompute.

\subsection{Inference, molecular dynamics, and the warm-start caveat}
\label{sec:ase}

Both models expose an ASE calculator~\cite{ase} returning energies, forces and
stress, with the absolute energy reconstructed from the $E_0$ values stored on
the checkpoint and, for a \texttt{regular} model, the recomputed baseline.
Warm-starting \OrbO's SCF and localizer from the previous dynamics step is
supported but off by default: a warm-started SCF converges to a density that
differs at the $10^{-9}$ level, which diagonalization within near-degenerate
subspaces and an iterative localizer amplify into a feature shift that puts a
model trained on cold features out of distribution, with nothing to flag it but
degraded accuracy. \OrbA\ runs no SCF or localizer and is deterministic and
history-independent by construction.

\section{Cost and complexity}
\label{sec:complexity}

For \OrbO, the per-frame featurization cost is the SCF, the localization and
the integral transformations, of which the $O(n_{\mathrm{MO}}^5)$ AO-to-MO
transformation of the two-electron integrals dominates asymptotically; the
one-electron transforms are $O(n_{\mathrm{MO}}^3)$. A small basis and window
pruning keep $n_{\mathrm{MO}}$ tractable, and the stage runs once, offline. For
\OrbA\ the overlap is $O(n_{\mathrm{AO}}^2)$, the two eigendecompositions are
$O(n_{\mathrm{AO}}^3)$ and form the bottleneck, and the contractions of
Eq.~\eqref{eq:blocksum} cost $O(n_{\mathrm{AO}}^2 n_{\mathrm{atoms}} +
n_{\mathrm{AO}} n_{\mathrm{atoms}}^2)$, on matrices a few tens on a side. There is no $O(n^5)$ term and no iteration count, but the solve is paid on every
forward pass. \OrbA's graph has one node per atom and \OrbO's one per retained
orbital, a small multiple of the atom count, which makes \OrbO's training step
the more expensive; at inference every \OrbO\ call needs an SCF and a
localization, which is also why a \texttt{regular} Hartree--Fock baseline is so
costly for \OrbA\ (Table~\ref{tab:cost}). Both models are implemented in
PyTorch on $e3nn$~\cite{e3nn} and the MACE~\cite{mace} modules; \OrbA's
electronic solve needs no external quantum-chemistry library, while \OrbO\ uses
PySCF~\cite{pyscf}.


\section{Supplementary results}
\label{si:results}

\subsection{The two models on revised MD17 at matched capacity}
\label{sec:rmd17-orbital}

On ANI-1x the two models differed in configuration, \OrbO\ running 2 layers at
$r_{\max}=6$\,{\AA} against \OrbA's 3 at $5$\,{\AA}. The rMD17 campaign removes
most of that confound: both models run at 4 layers and 150 channels with the
same correlation order, mean-squared loss and force-to-energy weight ratio,
optimizer, learning rate, batch size, seed and 12,000 epoch cap, on identical
splits, differing only in each model's default cutoff.
Table~\ref{tab:rmd17-matched} gives means over all ten molecules.

\begin{table}[!tbp]
\centering
\caption{Revised MD17 at matched depth and width, means over all ten molecules,
one model per molecule. Total energy MAE $E$ in meV, force MAE $F$ in
meV/{\AA}. The ratio columns give \OrbO\ over \OrbA.}
\label{tab:rmd17-matched}
\vspace{5pt}
\small 
\setlength{\tabcolsep}{2pt}
\begin{tabular}{lrrrrrr}
\toprule
 & \multicolumn{2}{c}{\OrbA} & \multicolumn{2}{c}{\OrbO} & \multicolumn{2}{c}{ratio} \\
\cmidrule(lr){2-3}\cmidrule(lr){4-5}\cmidrule(lr){6-7}
Learning scheme & $E$ & $F$ & $E$ & $F$ & $E$ & $F$ \\
\midrule
direct               & 2.37 & 5.1 & 66.42 & 29.5 & 28.1 & 5.7 \\
regular HF           & 1.41 & 3.7 & 44.21 & 17.8 & 31.4 & 4.8 \\
regular xTB          & \textbf{0.58} & \textbf{2.7} & \textbf{27.23} & \textbf{10.5} & 46.9 & 3.9 \\
\texttt{mfdh} HF     & 6.76 & 5.9 & 68.14 & 30.3 & 10.1 & 5.1 \\
\texttt{mfdh} xTB    & 4.87 & 4.9 & 66.71 & 29.9 & 13.7 & 6.1 \\
\bottomrule
\end{tabular}
\end{table}

Within \OrbO\ the regular schemes lead as they do within \OrbA, regular xTB at
$27.2$\,meV and regular HF at $44.2$, with the direct and \texttt{mfdh} arms
trailing together at $66.4$ to $68.1$, whereas \OrbA's \texttt{mfdh} arms are
markedly worse than its direct model. \OrbA\ is ahead in every arm, so removing
the depth and cutoff confound does not reverse the conclusion of
Sec.~\ref{sec:ani-analysis}.

The gap is larger than on ANI-1x, a factor of 28 in energy and 6 in force on
the direct arm against three there, and the two models also train differently.
Over all seventy runs in each family \OrbO\ reaches a median final training
loss of $0.24$ against $0.014$ and stops at a median epoch of $1{,}737$ against
$6{,}688$, its plateau scheduler having reached the $10^{-6}$ learning-rate
floor before the $12{,}000$ epoch cap that \OrbA\ reaches. It also carries
more parameters at the same width, because its nodes span five to ten orbital
species where \OrbA's span two to four elements, and through its equivariant
orbital embedding and
Mulliken-pooled atom head, so the comparison is matched in depth and width but
not in parameter count. On naphthalene \OrbO's regular-HF arm reaches
$80.8$\,meV against $46.6$ for its direct arm, the only case in either family
where an HF baseline substantially worsens the energy; dropping naphthalene
moves the two regular ratios only from $31.4$ to $30.9$ and from $46.9$ to
$46.6$.

\makeatletter
\setlength{\@dblfptop}{0pt}
\setlength{\@dblfpsep}{12pt plus 2pt}
\setlength{\@dblfpbot}{0pt plus 1fil}
\makeatother
\begin{table*}[!tbp]
\centering
\caption{The broadened form of Eq.~\eqref{eq:broadening} does not converge at a
degeneracy. Relative error of the analytic gradient against a double-precision
central finite difference, over the broadening parameter $\epsilon$, followed by
the change in that gradient under a random orthogonal rotation of the
eigenvectors within each degenerate block at $\epsilon = 10^{-12}$. Water is a
control with no degeneracy, where the error falls to the finite-difference floor
and the rotation has no block to act on; on the other three the error stops
improving below $\epsilon \approx 10^{-4}$ and settles at tens of percent, and
the surviving error is of the same size as the change produced by rotating the
basis, so it is not a function of the geometry alone.}
\label{tab:spectral-eps}
\vspace{5pt}
\footnotesize
\begin{tabular}{@{}lccccccc@{}}
\toprule
& \multicolumn{6}{c}{relative error at $\epsilon =$}
  & basis \\
\cmidrule(lr){2-7}
Molecule & $10^{-2}$ & $10^{-4}$ & $10^{-6}$ & $10^{-8}$ & $10^{-10}$
  & $10^{-12}$ & rotation \\
\midrule
H$_2$O (control)    & \num{1.8e-2} & \num{4.6e-4} & \num{4.7e-6} & \num{4.8e-8}
  & \num{1.4e-9} & \num{1.4e-9} & \num{6.1e-15} \\
CH$_4$ ($T_d$)      & \num{1.7e-1} & \num{1.0e-1} & \num{1.0e-1} & \num{1.0e-1}
  & \num{1.0e-1} & \num{1.0e-1} & \num{2.5e-1} \\
NH$_3$ ($C_{3v}$)   & \num{5.9e-1} & \num{2.8e-1} & \num{2.8e-1} & \num{2.8e-1}
  & \num{2.8e-1} & \num{2.8e-1} & \num{2.9e-1} \\
C$_6$H$_6$ ($D_{6h}$) & \num{4.4e-1} & \num{3.0e-1} & \num{2.9e-1} & \num{2.9e-1}
  & \num{2.9e-1} & \num{2.9e-1} & \num{3.1e-1} \\
\bottomrule
\end{tabular}
\end{table*}

\begin{table*}[!tbp]
\centering
\caption{Accuracy of the spectral derivative split by spectral function.
Relative error against a double-precision central finite difference, reported
for the energy-weighted density matrix $\mathbf{W}$ and, for reference, for
$\mathbf{P}$. Broadening is applied to the density-matrix step alone, which is
what isolates the error to $\mathbf{W}$. $\mathbf{P}$'s residual is not
identically zero: degeneracy contributes nothing to it, but the occupied/virtual
pairs carry the HOMO--LUMO gap in the denominator, where broadening perturbs
$1/d$ by $O(\epsilon/d^2)$ and the error duly vanishes with $\epsilon$. The
claim the last column supports is therefore the ratio rather than a zero.}
\label{tab:spectral}
\vspace{5pt}
\small
\begin{tabular}{@{}llccr@{}}
\toprule
Molecule & Scheme & rel.\ err.\ in $\mathbf{W}$ & rel.\ err.\ in $\mathbf{P}$
  & ratio \\
\midrule
CH$_4$   & divided difference (Eq.~\ref{eq:dk})        & \num{9.1e-10} & \num{5.5e-10}
  & $2$ \\
CH$_4$   & broadened, $\epsilon=10^{-6}$               & \num{7.6e-2}  & \num{2.1e-6}
  & $4\times 10^{4}$ \\
CH$_4$   & broadened, $\epsilon=10^{-12}$              & \num{7.6e-2}  & \num{5.4e-10}
  & $1\times 10^{8}$ \\
C$_6$H$_6$ & divided difference (Eq.~\ref{eq:dk})      & \num{5.7e-10} & \num{3.5e-10}
  & $2$ \\
C$_6$H$_6$ & broadened, $\epsilon=10^{-6}$             & \num{3.9e-2}  & \num{2.5e-5}
  & $2\times 10^{3}$ \\
\bottomrule
\end{tabular}
\end{table*}

\subsection{Validating the spectral derivative}

Section~\ref{sec:spectral} claimed that broadened eigenvector perturbation
theory is qualitatively wrong at degeneracies and that the error is specific to
$\mathbf{W}$. The analytic derivative of a random linear functional of the
features was compared against a double-precision central finite difference
under three schemes: the divided difference of Eq.~\eqref{eq:dk}, the
broadened form of Eq.~\eqref{eq:broadening} for $\epsilon$ from $10^{-2}$ to
$10^{-12}$, and the broadened form after a random orthogonal rotation of the
eigenvectors within each degenerate block. Methane's $t_2$ triplet, ammonia's
$e$ doublet and benzene's $\pi$ pairs are exactly degenerate, with smallest
occupied gaps of $10^{-15}\,\Eh$ or below; water, whose smallest occupied gap is
$2.3\times 10^{-2}\,\Eh$, is the control on which broadening must converge.
Geometries were built from shared trigonometric literals, since an ammonia
geometry given to four decimals is only $C_{3v}$ to about $3\times 10^{-4}$,
which opens the doublet to $6\times 10^{-5}$ and silently retires it from the
test. The divided difference agrees with the finite difference to between
$3\times 10^{-10}$ and $5\times 10^{-9}$ relative on all four molecules, the
floor of a double-precision central difference, and rotating the eigenvectors
within a degenerate block moves it by at most $9\times 10^{-15}$, confirming
Proposition~\ref{prop:basis} numerically.

Broadening fails as predicted (Table~\ref{tab:spectral-eps}). On the control its
error falls by four orders of magnitude to the finite-difference floor; on each
degenerate molecule it stops improving below $\epsilon \approx 10^{-4}$ and
settles at $10$ to $29$ percent, and rotating the eigenvectors within a block
changes the gradient by $0.25$ to $0.31$, comparable to the error itself. The
broadened derivative is therefore not a function of the geometry at all but of
the basis the eigensolver happened to return. The mechanism is a single
annihilated term. For an occupied pair sharing an occupation, $\mathbf{W}$'s
divided difference is
$(2 f_i \epsilon_i - 2 f_j \epsilon_j)/(\epsilon_i - \epsilon_j) = 2f$,
evaluated as $+2.000$ on an exactly degenerate pair, while broadening returns
$2 f d^2/(d^2+\epsilon) \to 0$, evaluated as $0.000$. Within a block this
leaves $\mathbf{L} = f'(\omega)\mathbf{I}$ in place of
$f'(\omega)\mathbf{1}\mathbf{1}^{\!\top}$, whose Hadamard product is not
invariant under rotation of the block. For $\mathbf{P}$ the numerator
$2(f_i - f_j)$ vanishes identically across a degenerate pair, and both schemes
return bit-identical $\mathbf{L}$.

\balance
Table~\ref{tab:spectral} confirms the localization to $\mathbf{W}$. Under
broadening $\mathbf{W}$ is wrong by about eight percent for methane and four for
benzene, independently of $\epsilon$, while $\mathbf{P}$'s error is three to
eight orders of magnitude smaller and falls with $\epsilon$, from
$2.1\times 10^{-6}$ to $5.4\times 10^{-10}$ as $\epsilon$ goes from $10^{-6}$ to
$10^{-12}$, with $\mathbf{W}$ fixed at $7.6\times 10^{-2}$. The localization
holds within the $\mathbf{P}/\mathbf{W}$ pair only: broadening the
orthogonalizer $\mathbf{S}^{-1/2}$ as well, whose spectrum is as degenerate as
the Hamiltonian's in a symmetric molecule, corrupts $\mathbf{P}$ by around
$2\times 10^{-1}$ upstream, which is why the $\mathbf{P}$ column applies
broadening to the density-matrix step alone.


\clearpage
\clearpage
\bibliographystyle{unsrt}
\bibliography{refs}

@article{ace,
  author  = {Drautz, Ralf},
  title   = {{Atomic cluster expansion for accurate and transferable interatomic potentials}},
  journal = {Phys. Rev. B},
  volume  = {99},
  pages   = {014104},
  year    = {2019}
}

@inproceedings{mace,
  author    = {Batatia, Ilyes and Kov\'acs, D\'avid P\'eter and Simm, Gregor N. C. and Ortner, Christoph and Cs\'anyi, G\'abor},
  title     = {{MACE: Higher Order Equivariant Message Passing Neural Networks for Fast and Accurate Force Fields}},
  booktitle = {Advances in Neural Information Processing Systems (NeurIPS)},
  year      = {2022}
}

@article{bpnn,
  author  = {Behler, J\"org and Parrinello, Michele},
  title   = {{Generalized Neural-Network Representation of High-Dimensional Potential-Energy Surfaces}},
  journal = {Phys. Rev. Lett.},
  volume  = {98},
  pages   = {146401},
  year    = {2007},
  doi     = {10.1103/PhysRevLett.98.146401}
}

@article{soap,
  author  = {Bart\'ok, Albert P. and Kondor, Risi and Cs\'anyi, G\'abor},
  title   = {{On representing chemical environments}},
  journal = {Phys. Rev. B},
  volume  = {87},
  pages   = {184115},
  year    = {2013},
  doi     = {10.1103/PhysRevB.87.184115}
}

@article{nequip,
  author  = {Batzner, Simon and Musaelian, Albert and Sun, Lixin and Geiger, Mario and Mailoa, Jonathan P. and Kornbluth, Mordechai and Molinari, Nicola and Smidt, Tess E. and Kozinsky, Boris},
  title   = {{E(3)-equivariant graph neural networks for data-efficient and accurate interatomic potentials}},
  journal = {Nat. Commun.},
  volume  = {13},
  pages   = {2453},
  year    = {2022},
  doi     = {10.1038/s41467-022-29939-5}
}

@misc{e3nn,
  author       = {Geiger, Mario and Smidt, Tess and others},
  title        = {{e3nn: Euclidean Neural Networks}},
  howpublished = {\url{https://github.com/e3nn/e3nn}},
  year         = {2022}
}

@inproceedings{dimenet,
  author    = {Gasteiger, Johannes and Gro{\ss}, Janek and G\"unnemann, Stephan},
  note      = {Published under the name Klicpera},
  title     = {{Directional Message Passing for Molecular Graphs}},
  booktitle = {International Conference on Learning Representations (ICLR)},
  year      = {2020}
}

@article{orbnet,
  author  = {Qiao, Zhuoran and Welborn, Matthew and Anandkumar, Animashree and Manby, Frederick R. and {Miller III}, Thomas F.},
  title   = {{OrbNet: Deep learning for quantum chemistry using symmetry-adapted atomic-orbital features}},
  journal = {J. Chem. Phys.},
  volume  = {153},
  pages   = {124111},
  year    = {2020},
  doi     = {10.1063/5.0021955}
}

@inproceedings{darkforces,
  author    = {Bigi, Filippo and Langer, Marcel F. and Ceriotti, Michele},
  title     = {{The dark side of the forces: assessing non-conservative force models for atomistic machine learning}},
  booktitle = {Proceedings of the 42nd International Conference on Machine Learning (ICML)},
  series    = {Proceedings of Machine Learning Research},
  volume    = {267},
  pages     = {4384--4414},
  year      = {2025},
  doi       = {10.48550/arXiv.2412.11569}
}

@article{pyscf,
  author  = {Sun, Qiming and others},
  title   = {{PySCF: The Python-based Simulations of Chemistry Framework}},
  journal = {WIREs Comput. Mol. Sci.},
  volume  = {8},
  pages   = {e1340},
  year    = {2018}
}

@article{boys,
  author  = {Boys, S. F.},
  title   = {{Construction of Some Molecular Orbitals to Be Approximately Invariant for Changes from One Molecule to Another}},
  journal = {Rev. Mod. Phys.},
  volume  = {32},
  pages   = {296},
  year    = {1960}
}

@article{pm,
  author  = {Pipek, J\'anos and Mezey, Paul G.},
  title   = {{A fast intrinsic localization procedure applicable for ab initio and semiempirical linear combination of atomic orbital wave functions}},
  journal = {J. Chem. Phys.},
  volume  = {90},
  pages   = {4916},
  year    = {1989}
}

@article{mulliken,
  author  = {Mulliken, R. S.},
  title   = {{Electronic Population Analysis on LCAO-MO Molecular Wave Functions. I}},
  journal = {J. Chem. Phys.},
  volume  = {23},
  pages   = {1833},
  year    = {1955}
}

@article{wh,
  author  = {Wolfsberg, Max and Helmholz, Lindsay},
  title   = {{The Spectra and Electronic Structure of the Tetrahedral Ions MnO$_4^-$, CrO$_4^{--}$, and ClO$_4^-$}},
  journal = {J. Chem. Phys.},
  volume  = {20},
  pages   = {837},
  year    = {1952}
}

@article{hoffmann,
  author  = {Hoffmann, Roald},
  title   = {{An Extended H\"uckel Theory. I. Hydrocarbons}},
  journal = {J. Chem. Phys.},
  volume  = {39},
  pages   = {1397},
  year    = {1963}
}

@article{lowdin,
  author  = {L\"owdin, Per-Olov},
  title   = {{On the Non-Orthogonality Problem Connected with the Use of Atomic Wave Functions in the Theory of Molecules and Crystals}},
  journal = {J. Chem. Phys.},
  volume  = {18},
  pages   = {365},
  year    = {1950}
}

@article{mayer,
  author  = {Mayer, I.},
  title   = {{Charge, bond order and valence in the ab initio SCF theory}},
  journal = {Chem. Phys. Lett.},
  volume  = {97},
  pages   = {270},
  year    = {1983}
}

@article{hehre,
  author  = {Hehre, W. J. and Stewart, R. F. and Pople, J. A.},
  title   = {{Self-Consistent Molecular-Orbital Methods. I. Use of Gaussian Expansions of Slater-Type Atomic Orbitals}},
  journal = {J. Chem. Phys.},
  volume  = {51},
  pages   = {2657},
  year    = {1969}
}

@article{pople631g,
  author  = {Hariharan, P. C. and Pople, John A.},
  title   = {{The influence of polarization functions on molecular orbital hydrogenation energies}},
  journal = {Theor. Chim. Acta},
  volume  = {28},
  pages   = {213--222},
  year    = {1973},
  doi     = {10.1007/BF00533485}
}

@article{wb97x,
  author  = {Chai, Jeng-Da and Head-Gordon, Martin},
  title   = {{Systematic optimization of long-range corrected hybrid density functionals}},
  journal = {J. Chem. Phys.},
  volume  = {128},
  pages   = {084106},
  year    = {2008},
  doi     = {10.1063/1.2834918}
}

@article{mcmurchie,
  author  = {McMurchie, Larry E. and Davidson, Ernest R.},
  title   = {{One- and two-electron integrals over Cartesian Gaussian functions}},
  journal = {J. Comput. Phys.},
  volume  = {26},
  pages   = {218},
  year    = {1978}
}

@article{gfn2,
  author  = {Bannwarth, Christoph and Ehlert, Sebastian and Grimme, Stefan},
  title   = {{GFN2-xTB: An Accurate and Broadly Parametrized Self-Consistent Tight-Binding Quantum Chemical Method}},
  journal = {J. Chem. Theory Comput.},
  volume  = {15},
  pages   = {1652},
  year    = {2019}
}

@book{zbl,
  author    = {Ziegler, J. F. and Biersack, J. P. and Littmark, U.},
  title     = {{The Stopping and Range of Ions in Solids}},
  publisher = {Pergamon},
  address   = {New York},
  year      = {1985}
}

@article{daleckii,
  author  = {Daleckii, Ju. L. and Krein, S. G.},
  title   = {{Integration and differentiation of functions of Hermitian operators and applications to the theory of perturbations}},
  journal = {Amer. Math. Soc. Transl. (2)},
  volume  = {47},
  pages   = {1},
  year    = {1965}
}

@book{bhatia,
  author    = {Bhatia, Rajendra},
  title     = {{Matrix Analysis}},
  publisher = {Springer},
  address   = {New York},
  year      = {1997},
  note      = {Ch.~V}
}

@book{higham,
  author    = {Higham, Nicholas J.},
  title     = {{Functions of Matrices: Theory and Computation}},
  publisher = {SIAM},
  address   = {Philadelphia},
  year      = {2008}
}

@article{ewald,
  author  = {Ewald, P. P.},
  title   = {{Die Berechnung optischer und elektrostatischer Gitterpotentiale}},
  journal = {Ann. Phys.},
  volume  = {369},
  pages   = {253},
  year    = {1921}
}

@article{deltaml,
  author  = {Ramakrishnan, Raghunathan and Dral, Pavlo O. and Rupp, Matthias and von Lilienfeld, O. Anatole},
  title   = {{Big Data Meets Quantum Chemistry Approximations: The $\Delta$-Machine Learning Approach}},
  journal = {J. Chem. Theory Comput.},
  volume  = {11},
  pages   = {2087--2096},
  year    = {2015},
  doi     = {10.1021/acs.jctc.5b00099}
}

@article{ema,
  author  = {Polyak, B. T. and Juditsky, A. B.},
  title   = {{Acceleration of Stochastic Approximation by Averaging}},
  journal = {SIAM J. Control Optim.},
  volume  = {30},
  pages   = {838},
  year    = {1992}
}

@inproceedings{swa,
  author    = {Izmailov, Pavel and Podoprikhin, Dmitrii and Garipov, Timur and Vetrov, Dmitry and Wilson, Andrew Gordon},
  title     = {{Averaging Weights Leads to Wider Optima and Better Generalization}},
  booktitle = {Conference on Uncertainty in Artificial Intelligence (UAI)},
  year      = {2018}
}

@article{huber,
  author  = {Huber, Peter J.},
  title   = {{Robust Estimation of a Location Parameter}},
  journal = {Ann. Math. Statist.},
  volume  = {35},
  pages   = {73},
  year    = {1964}
}

@article{ase,
  author  = {Larsen, Ask Hjorth and others},
  title   = {{The atomic simulation environment: a Python library for working with atoms}},
  journal = {J. Phys.: Condens. Matter},
  volume  = {29},
  pages   = {273002},
  year    = {2017}
}

@article{ani1x,
  author  = {Smith, Justin S. and Nebgen, Ben and Lubbers, Nicholas and Isayev, Olexandr and Roitberg, Adrian E.},
  title   = {{Less is more: Sampling chemical space with active learning}},
  journal = {J. Chem. Phys.},
  volume  = {148},
  pages   = {241733},
  year    = {2018},
  doi     = {10.1063/1.5023802}
}

@article{md17,
  author  = {Chmiela, Stefan and Tkatchenko, Alexandre and Sauceda, Huziel E. and Poltavsky, Igor and Sch\"utt, Kristof T. and M\"uller, Klaus-Robert},
  title   = {{Machine learning of accurate energy-conserving molecular force fields}},
  journal = {Sci. Adv.},
  volume  = {3},
  pages   = {e1603015},
  year    = {2017},
  doi     = {10.1126/sciadv.1603015}
}

@article{rmd17,
  author  = {Christensen, Anders S. and von Lilienfeld, O. Anatole},
  title   = {{On the role of gradients for machine learning of molecular energies and forces}},
  journal = {Mach. Learn.: Sci. Technol.},
  volume  = {1},
  pages   = {045018},
  year    = {2020},
  doi     = {10.1088/2632-2153/abba6f}
}

@article{threebpa,
  author  = {Kov\'acs, D\'avid P\'eter and van der Oord, Cas and Kucera, Jiri and Allen, Alice E. A. and Cole, Daniel J. and Ortner, Christoph and Cs\'anyi, G\'abor},
  title   = {{Linear Atomic Cluster Expansion Force Fields for Organic Molecules: Beyond RMSE}},
  journal = {J. Chem. Theory Comput.},
  volume  = {17},
  pages   = {7696--7711},
  year    = {2021},
  doi     = {10.1021/acs.jctc.1c00647}
}

@article{orbnetdenali,
  author  = {Christensen, Anders S. and Sirumalla, Sai Krishna and Qiao, Zhuoran and O'Connor, Michael B. and Smith, Daniel G. A. and Ding, Feizhi and Bygrave, Peter J. and Anandkumar, Animashree and Welborn, Matthew and Manby, Frederick R. and {Miller III}, Thomas F.},
  title   = {{OrbNet Denali: A machine learning potential for biological and organic chemistry with semi-empirical cost and DFT accuracy}},
  journal = {J. Chem. Phys.},
  volume  = {155},
  pages   = {204103},
  year    = {2021},
  doi     = {10.1063/5.0061990}
}

@article{orbnetequi,
  author  = {Qiao, Zhuoran and Christensen, Anders S. and Welborn, Matthew and Manby, Frederick R. and Anandkumar, Anima and {Miller III}, Thomas F.},
  title   = {{Informing geometric deep learning with electronic interactions to accelerate quantum chemistry}},
  journal = {Proc. Natl. Acad. Sci. U.S.A.},
  volume  = {119},
  pages   = {e2205221119},
  year    = {2022},
  doi     = {10.1073/pnas.2205221119}
}

@article{schnorb,
  author  = {Sch\"utt, K. T. and Gastegger, M. and Tkatchenko, A. and M\"uller, K.-R. and Maurer, R. J.},
  title   = {{Unifying machine learning and quantum chemistry with a deep neural network for molecular wavefunctions}},
  journal = {Nat. Commun.},
  volume  = {10},
  pages   = {5024},
  year    = {2019},
  doi     = {10.1038/s41467-019-12875-2}
}

@inproceedings{phisnet,
  author    = {Unke, Oliver and Bogojeski, Mihail and Gastegger, Michael and Geiger, Mario and Smidt, Tess and M\"uller, Klaus-Robert},
  title     = {{SE(3)-equivariant prediction of molecular wavefunctions and electronic densities}},
  booktitle = {Advances in Neural Information Processing Systems 34 (NeurIPS)},
  year      = {2021},
  doi       = {10.48550/arXiv.2106.02347}
}

@article{spookynet,
  author  = {Unke, Oliver T. and Chmiela, Stefan and Gastegger, Michael and Sch\"utt, Kristof T. and Sauceda, Huziel E. and M\"uller, Klaus-Robert},
  title   = {{SpookyNet: Learning force fields with electronic degrees of freedom and nonlocal effects}},
  journal = {Nat. Commun.},
  volume  = {12},
  pages   = {7273},
  year    = {2021},
  doi     = {10.1038/s41467-021-27504-0}
}

@article{aimnet2,
  author  = {Anstine, Dylan M. and Zubatyuk, Roman and Isayev, Olexandr},
  title   = {{AIMNet2: a neural network potential to meet your neutral, charged, organic, and elemental-organic needs}},
  journal = {Chem. Sci.},
  volume  = {16},
  number  = {23},
  pages   = {10228--10244},
  year    = {2025},
  doi     = {10.1039/d4sc08572h}
}

\end{document}